\documentclass[reprint,amsmath,amssymb,aip,jcp]{revtex4-2}

\usepackage{graphicx} % Include figure files
\usepackage{bmpsize}
\usepackage{dcolumn} % Align table columns on decimal point
\usepackage{bm} % bold math
\usepackage{here}
\usepackage{color}
\usepackage[utf8]{inputenc}
\usepackage[T1]{fontenc}
\usepackage{mathptmx}
\usepackage{etoolbox}
\usepackage[english]{babel}

\makeatletter
\def\@email#1#2{%
 \endgroup
 \patchcmd{\titleblock@produce}
  {\frontmatter@RRAPformat}
  {\frontmatter@RRAPformat{\produce@RRAP{*#1\href{mailto:#2}{#2}}}\frontmatter@RRAPformat}
  {}{}
}%
\makeatother

\begin{document}
%%%%%%%%%%%%%%%%%%%%%%%%%%%%%%%%%%%%%%%%%%%%%%%%%%%%%%%%%%%%%%%%%%%%%%%%

%\preprint{AIP/123-QED}

%%%%%%%%%%%%%%%%%%%%%%%%%%%%%%%%%%%%%%%%%%%%%%%%%%%%%%%%%%%%%%%%%%%%%%%%
\title{Computation of the heat capacity of water from first principles}
%%%%%%%%%%%%%%%%%%%%%%%%%%%%%%%%%%%%%%%%%%%%%%%%%%%%%%%%%%%%%%%%%%%%%%%%

\author{Motoyuki Shiga}
\email{shiga.motoyuki@jaea.go.jp}
\affiliation{Center for Computational Science and e-Systems, Japan Atomic Energy Agency, Kashiwa, Chiba 277-0871, Japan}

\author{Jan Elsner}
\affiliation{Lehrstuhl f\"ur Theoretische Chemie II, Ruhr-Universit\"at Bochum, 44780 Bochum, Germany}
\affiliation{Research Center Chemical Sciences and Sustainability, Research Alliance Ruhr, 44780 Bochum, Germany}

\author{J\"org Behler}
\affiliation{Lehrstuhl f\"ur Theoretische Chemie II, Ruhr-Universit\"at Bochum, 44780 Bochum, Germany}
\affiliation{Research Center Chemical Sciences and Sustainability, Research Alliance Ruhr, 44780 Bochum, Germany}

\author{Bo Thomsen}
%\email{thomsen.bo@jaea.go.jp}
\affiliation{Center for Computational Science and e-Systems, Japan Atomic Energy Agency, Kashiwa, Chiba 277-0871, Japan}

\date{\today}

%%%%%%%%%%%%%%%%%%%%%%%%%%%%%%%%%%%%%%%%%%%%%%%%%%%%%%%%%%%%%%%%%%%%%%%%
\begin{abstract}
%%%%%%%%%%%%%%%%%%%%%%%%%%%%%%%%%%%%%%%%%%%%%%%%%%%%%%%%%%%%%%%%%%%%%%%%

Water is a unique solvent with many remarkable properties. An example is its exceptionally high heat capacity, which plays an important role in storing and transporting thermal energy, with implications for many processes from regulating the body temperature of living organisms to moderating our climate at the global scale. To elucidate the microscopic origin of the heat capacity of water from first principles, highly accurate computer simulations are required. Apart from a reliable description of the atomic interactions, the presence of light hydrogen atoms necessitates the explicit consideration of nuclear quantum effects through path integral molecular dynamics (PIMD) simulations. The high computational costs of PIMD simulations, which are even further increased by the need for an extensive statistical sampling of energy fluctuations to determine the heat capacity, can be strongly reduced by replacing first principles calculations with machine learning potentials to represent the atomic interactions. In this study, we use high-dimensional neural network potentials (HDNNPs) constructed from density functional theory calculations employing the RPBE-D3 and revPBE0-D3 functionals. To further enhance the computational performance, we introduce a highly efficient PIMD algorithm computing in parallel not only the energies and forces but also the coordinate and thermostat time evolutions. Using this approach, we are able to determine converged data for the heat capacity from a 4 ns simulation employing 128 beads. In particular, for the revPBE0-D3 functional we find excellent agreement with experiment, providing evidence that our approach represents a promising framework for the quantitative understanding of the thermodynamic properties of water and aqueous solutions.

%%%%%%%%%%%%%%%%%%%%%%%%%%%%%%%%%%%%%%%%%%%%%%%%%%%%%%%%%%%%%%%%%%%%%%%%
\end{abstract}
%%%%%%%%%%%%%%%%%%%%%%%%%%%%%%%%%%%%%%%%%%%%%%%%%%%%%%%%%%%%%%%%%%%%%%%%

%%%%%%%%%%%%%%%%%%%%%%%%%%%%%%%%%%%%%%%%%%%%%%%%%%%%%%%%%%%%%%%%%%%%%%%%
\maketitle
%%%%%%%%%%%%%%%%%%%%%%%%%%%%%%%%%%%%%%%%%%%%%%%%%%%%%%%%%%%%%%%%%%%%%%%%

\section{Introduction}

Liquid water has an exceptionally high heat capacity, making it possible to efficiently absorb, store, and release large amounts of thermal energy. \cite{eisenberg1969structure} This unique property plays a vital role in a wide range of fields, including climate science, biological processes, and industrial applications such as cooling systems and thermal energy storage. To understand why the heat capacity of water is so high, it is essential to reproduce and predict this property accurately from theoretical calculations.
{Thus, ab initio approach to thermodynamics of liquid water has been an important topic.
\cite{bukowski2007predictions}}

{}
At the atomistic level, two main factors are believed
to contribute to the heat capacity of water: the flexibility of its
hydrogen-bond network and the quantum nature of its atomic nuclei.
Upon heating, water molecules are vibrationally excited, and hydrogen bonds deform or break, thereby absorbing energy. The flexibility of the hydrogen bonds allows them to function as a thermal reservoir, which increases the heat capacity.
In contrast, internal vibrations and librations of water molecules involve high-frequency motions of the light hydrogen atoms, where quantum effects become significant. These effects cause vibrational energy levels to be discrete rather than continuous, which reduces the system’s ability to absorb energy efficiently. Consequently, the quantum nature of the atomic nuclei decreases the heat capacity, and in theoretical modeling it is crucial to achieve an appropriate balance between these two competing effects, which both need to be described with high accuracy.

Numerous studies have been conducted to calculate the heat capacity of water using classical force fields (CFFs), \cite{shinoda2005quantum,shiga2005calculation,paesani2006accurate,ceriotti2011accelerating,shvab2016atomistic,eltareb2021nuclear}. Early approaches employed classical molecular dynamics (MD) simulations with intramolecular vibrations of water molecules constrained. However, due to the lack of quantum effects, these methods were unable to provide quantitatively accurate results. It was not until 2005 that quantum effects were incorporated
{to compute the heat capacity of water}\cite{shinoda2005quantum,shiga2005calculation}
 via path integral molecular dynamics (PIMD).
\cite{parrinello1984study,hall1984nonergodicity,tuckerman1993efficient}
Using the non-polarizable SPC/F2 water model, this study successfully reproduced the heat capacity of ice and water vapor, but underestimated that of liquid water by approximately 20\%. This discrepancy was attributed to the insufficient representation of fluctuations in hydrogen bond energies in the model. Accurately capturing the quantum effects of intramolecular vibrations requires about one hundred beads (imaginary time slices) in PIMD, which significantly increases the computational cost compared to classical MD. Subsequently, an improved non-polarizable model, TIP4PQ/2005, was used in path integral Monte Carlo (PIMC) simulations,\cite{vega2010heat,noya2011path} yielding better agreement with experimental values. In this model, intramolecular vibrations were constrained, and their quantum contributions were omitted, focusing solely on quantum effects in intermolecular degrees of freedom and effectively reducing the number of beads required. Recently, quantum effects were predicted using phenomenological models based on phonon calculations using non-polarizable \cite{berta2020nuclear} or polarizable \cite{savoia2025influence} force fields. All previous studies have been based on CFFs, and no attempt has yet been made to evaluate the heat capacity using first-principles calculations that fully incorporate quantum effects with respect to all degrees of freedom.

The objective of this study is to fill this gap and to compute the heat capacity of water from first principles by PIMD simulations based on fundamental physical laws. Specifically, our aim is to derive the heat capacity by simulating the detailed atomic-scale dynamics of water and analyzing the resulting energy fluctuations.
To accomplish this, two key technical requirements must be satisfied. First, a highly accurate atomic interaction potential is needed to capture the flexibility of the hydrogen-bond network faithfully. Second, nuclear quantum effects within the complex many-body system must be incorporated into the simulations.

To meet both of these requirements simultaneously, we represent the Born-Oppenheimer energies with first-principles quality using high-dimensional neural network potentials (HDNNPs), \cite{behler2007generalized,behler2011neural,behler2015constructing,behler2016perspective,behler2021four} enabling an approach consistent with ab initio PIMD. \cite{marx1996ab,shiga2001unified} Specifically, we used the HDNNP trained on first-principles data obtained from density functional theory (DFT) calculations using the RPBE and revPBE0 exchange-correlation functionals with D3 dispersion corrections. These potentials naturally account for complex electronic effects such as intermolecular interactions, charge transfer, and molecular polarization, which are often neglected in conventional non-polarizable CFFs.
In recent years, the combination of PIMD and HDNNPs has been increasingly employed as an effective approach to accurately describe the structure and thermodynamics of liquid water. \cite{ChengHDNNP2016,P4971,P5631,ChengHDNNP2019,ReinhardtHDNNP2021,daru2022coupled,thomsen2024self,stolte2024nuclear,stolte2025random,malik2025accurate}
For a more thorough review of machine learning methods for water and aqueous systems, we refer to the recent review by Omranpour \textit{et al}.\cite{OmranpourReview}

Because the heat capacity is derived from fluctuations in the total energy, PIMD simulations typically need to run for several nanoseconds to achieve statistical convergence. Moreover, at ambient temperature, around one hundred beads are required, making the high computational cost of PIMD a significant challenge.
To address this problem, we have developed and implemented a new parallel PIMD algorithm with high scalability.
In this approach (hereafter referred to as the `local scheme'), we efficiently perform bead-wise parallel computation of interatomic interactions, similar to the previous method by Ruiz \textit{et al.},\cite{ruiz2016hierarchical} hereafter referred to as the `shared scheme.'

In the shared scheme, all coordinate and velocity variables are shared across all processors, and their time evolution is computed locally by each processor.
In contrast, the local scheme improves efficiency by assigning variables locally to each processor and computing the time evolution only for those locally allocated variables.
Along with this modification, we optimize the transformation between Cartesian and normal mode coordinate representations, thereby minimizing communication overhead.
These improvements enable long-time PIMD simulations of large systems containing thousands of atoms.

Using this computational framework, which accurately captures both the quantum behavior of atomic nuclei and the complex structural fluctuations of the hydrogen-bond network, we demonstrate that the experimental heat capacity of liquid water under ambient conditions can be obtained in excellent agreement with experiment. This result highlights the importance of incorporating nuclear quantum effects in simulations of the thermodynamic properties of aqueous systems.

\section{Method}

\subsection{Equations of motion}

The equations of motion for PIMD, based on the combination of the normal mode representation \cite{cao1996adiabatic} and massive Nos\'e–Hoover chain (NHC) thermostats, \cite{nose1984unified,hoover1985canonical,martyna1992nose} are well-established.\cite{marx2009ab,tuckerman2010statistical,shiga2018path} They are presented here to provide the necessary background for the computational algorithm described in Subsection II B.

The PIMD method generates the quantum distribution of $N$ distinguishable atoms by performing classical canonical sampling in an extended configuration space consisting of $P$ replicas (beads).
Each bead $s$ ($1 \le s \le P$) corresponds to a discrete slice along imaginary time in the interval $[0, \beta\hbar]$, where $\beta = 1/k_{\rm B}T$, with the Boltzmann constant $k_{\rm B}$ and the temperature $T$.

%%%

Let $\left({R}_{I,x}^{(s)}, {R}_{I,y}^{(s)}, {R}_{I,z}^{(s)}\right)$ be the Cartesian coordinates of atom $I$ in bead $s$. The effective potential energy $V_{\rm eff}$ is defined as
\begin{gather}
V_{\rm eff} = \sum_{\alpha=x,y,z} \sum_{I=1}^N \sum_{s=1}^P \frac{M_I^{} P}{2\beta^2\hbar^2} \left( {R}_{I,\alpha}^{(s)} - {R}_{I,\alpha}^{(s-1)} \right)^2
\nonumber \\
+ \frac{1}{P} \sum_{s=1}^P V\left( {R}_{1,x}^{(s)},\cdots,{R}_{N,z}^{(s)} \right),
\label{eq:01}
\end{gather}
where $V$ is the physical interaction potential among the $N$ atoms, and $M_I^{}$ is the mass of atom $I$.
To improve computational efficiency, a normal mode transformation is applied to diagonalize the first term of Eq.~(\ref{eq:01}). This transformation is defined as
\begin{equation}
{Q}_{I,\alpha}^{(k)} = \frac{1}{\sqrt{P}} \sum_{s=1}^P U_{s,k}^{} {R}_{I,\alpha}^{(s)},
\quad
{R}_{I,\alpha}^{(s)} = {\sqrt{P}} \sum_{k=1}^P U_{s,k}^{} {Q}_{I,\alpha}^{(k)},
\label{eq:02}
\end{equation}
using the $k$-th eigenvector $(U_{1,k}^{},\cdots,U_{P,k}^{})$. 
The corresponding eigenvalue is given as $\lambda_k^{}$ for later purposes.
The first normal mode corresponds to the centroid of all beads, and is given by
\begin{equation}
{Q}_{I,\alpha}^{(1)} = \frac{1}{P} \sum_{s=1}^P {R}_{I,\alpha}^{(s)}
\label{eq:03}
\end{equation}
since $U_{s,1}^{} = \sqrt{1/P}$ for all $s$.

%%%

To ensure thermal equilibrium in PIMD simulations, it is necessary to properly apply a thermostat. One effective approach is the use of massive NHC thermostats.
In this method, for each normal mode $k$ of atom $I$, the time evolution of the position is given by
\begin{equation}
\dot{Q}_{I,\alpha}^{(k)} = \frac{P_{I,\alpha}^{(k)}}{\mu_I^{(k)}},
\label{eq:04}
\end{equation}
and the time evolution of the corresponding momentum is governed by
\begin{equation}
\dot{P}_{I,\alpha}^{(k)} = - \frac{\partial V_{\rm eff}}{\partial Q_{I,\alpha}^{(k)}} - P_{I,\alpha}^{(k)} \frac{p_{I,\alpha,1}^{(k)}}{m_1^{(k)}}.
\label{eq:05}
\end{equation}
For each thermostat chain of length $L$ ($1 \le j \le L$), the time evolution of the position and momentum of the thermostat variables is described as
\begin{equation}
\dot{\eta}_{I,\alpha,j}^{(k)} = \frac{p_{I,\alpha,j}^{(k)}}{m_j^{(k)}}
\label{eq:06}
\end{equation}
and
\begin{gather}
\dot{p}_{I,\alpha,j}^{(k)} = \frac{\left( P_{I,\alpha}^{(k)} \right)^2}{\mu_I^{(k)}} \delta_{j,1} + \frac{\left( p_{I,\alpha,j-1}^{(k)} \right)^2}{m_{j-1}^{(k)}} (1 - \delta_{j,1})
\nonumber \\
- \beta^{-1} - p_{I,\alpha,j}^{(k)} \frac{p_{I,\alpha,j+1}^{(k)}}{m_{j+1}^{(k)}} (1 - \delta_{j,L}),
\label{eq:07}
\end{gather}
respectively, where $\delta_{ij}^{}$ denotes the Kronecker delta.
The fictitious masses $\mu_I^{(k)}$ for the atoms and $m_j^{(k)}$ for the thermostats are tunable parameters in PIMD.
Typically, $\mu_I^{(1)}$ is set equal to the physical mass $M_I^{}$, while the other masses $\mu_I^{(k)}$ ($k \ne 1$) and $m_j^{(k)}$ are chosen to optimize stability and efficiency. Specific values depend on the system and implementation.
{
In this study, we have chosen as follows: 
$\mu_I^{(1)}=M_I^{}$,
$\mu_I^{(k)} = \frac{M_I^{}P\lambda_k^{}}{\beta^2\hbar^2}$ for $k \ne 1$,
$m_j^{(1)}=\frac{\tau_{\rm sys}^2}{\beta(2\pi)^2}$ with $\tau_{\rm sys}^{}=$ 100 fs, and
$m_j^{(k)}=\frac{\beta\hbar^2}{P}$
for $k \ne 1$.
}

%%%

The total energy of the system, defined as the PIMD pseudo-Hamiltonian
\begin{gather}
H_{\rm pimd} = \sum_{\alpha=x,y,z} \sum_{I=1}^N \left[ \frac{{P_{I,\alpha}^{(1)}}^2}{2M_I^{}} + \sum_{k=2}^P \frac{{P_{I,\alpha}^{(k)}}^2}{2\mu_I^{(k)}}
\right. \nonumber \\ \left.
+ \sum_{j=1}^L \left\{ \frac{{p_{I,\alpha,j}^{(k)}}^2}{2m_j^{(k)}} + \beta^{-1} \eta_{I,\alpha,j}^{(k)} \right\} \right] + V_{\rm eff}(\{{\bf Q}\}),
\label{eq:08}
\end{gather}
is conserved by the set of Eqs.~(\ref{eq:04})--(\ref{eq:07}).

%%%

Using the extended phase space representation
\[
\boldsymbol{\Gamma} = \{ \mathbf{Q}, \mathbf{P}, \boldsymbol{\eta}, \mathbf{p} \},
\]
the set of equations can be compactly written as
\begin{equation}
\frac{d\boldsymbol{\Gamma}}{dt} = i{\cal L} \boldsymbol{\Gamma},
\label{eq:09}
\end{equation}
where \({\cal L}\) is the Liouville operator, expressed as a sum of individual contributions,
\begin{equation}
{\cal L} = {\cal L}_{\rm A} + {\cal L}_{\rm B} + {\cal L}_{\rm C} + \sum_{k=1}^P {\cal L}_{\rm D}^{(k)}.
\label{eq:10}
\end{equation}
The individual terms are defined as
\begin{equation}
i{\cal L}_{\rm A} = \sum_{\alpha = x, y, z} \sum_{I=1}^N \sum_{k=1}^P 
\frac{P_{I,\alpha}^{(k)}}{\mu_I^{(k)}} \frac{\partial}{\partial Q_{I,\alpha}^{(k)}},
\label{eq:11}
\end{equation}
\begin{equation}
i{\cal L}_{\rm B} = - \sum_{\alpha = x, y, z} \sum_{k=1}^P \sum_{I=1}^N 
\frac{M_I^{} P \lambda^{(k)}}{\beta^2 \hbar^2} Q_{I,\alpha}^{(k)} \frac{\partial}{\partial P_{I,\alpha}^{(k)}},
\label{eq:12}
\end{equation}
\begin{equation}
i{\cal L}_{\rm C} = \sum_{\alpha = x, y, z} \sum_{k=1}^P \sum_{I=1}^N 
F_{I,\alpha}^{(k)} \frac{\partial}{\partial P_{I,\alpha}^{(k)}},
\label{eq:13}
\end{equation}
\begin{gather}
i{\cal L}_{\rm D}^{(k)} =
- \sum_{\alpha = x, y, z} \sum_{I=1}^N 
P_{I,\alpha}^{(k)} \frac{p_{I,\alpha,1}^{(k)}}{m_1^{(k)}} \frac{\partial}{\partial P_{I,\alpha}^{(k)}} \nonumber \\
+ \sum_{\alpha = x, y, z} \sum_{I=1}^N \sum_{j=1}^L \Bigg[
\frac{p_{I,\alpha,j}^{(k)}}{m_j^{(k)}} \frac{\partial}{\partial \eta_{I,\alpha,j}^{(k)}}
 \nonumber \\
+ \left\{
\frac{(P_{I,\alpha}^{(k)})^2}{\mu_I^{(k)}} \delta_{j,1}
+ \frac{(p_{I,\alpha,j-1}^{(k)})^2}{m_{j-1}^{(k)}} (1 - \delta_{j,1}^{}) \right. \nonumber \\
\left. - \beta^{-1}
- p_{I,\alpha,j}^{(k)} \frac{p_{I,\alpha,j+1}^{(k)}}{m_{j+1}^{(k)}} (1 - \delta_{j,L}^{})
\right\} \frac{\partial}{\partial p_{I,\alpha,j}^{(k)}}
\Bigg],
\label{eq:14}
\end{gather}
and the force in the normal mode representation is defined as
\begin{equation}
F_{I,\alpha}^{(k)} = -\frac{1}{\sqrt{P}} \sum_{s=1}^P U_{s,k} \frac{\partial V}{\partial R_{I,\alpha}^{(s)}}.
\label{eq:15}
\end{equation}
The time evolution governed by this Liouville operator can be solved numerically using the reference system propagator algorithm (RESPA). \cite{tuckerman1992reversible} The time evolution operator is approximately decomposed via symmetric Trotter splitting as
\begin{gather}
e^{i{\cal L} \Delta t} \approx\ 
e^{i{\cal L}_{\rm D}^{(1)} \frac{\Delta t}{2}} 
e^{i{\cal L}_{\rm C} \frac{\Delta t}{2}} \left[
e^{i \sum_{k=2}^L {\cal L}_{\rm D}^{(k)} \frac{\Delta t}{2n}} 
e^{i {\cal L}_{\rm B}^{(k)} \frac{\Delta t}{2n}} 
\right. \nonumber \\ \left. \times
e^{i {\cal L}_{\rm A}^{(k)} \frac{\Delta t}{n}} 
e^{i {\cal L}_{\rm B}^{(k)} \frac{\Delta t}{2n}} 
e^{i \sum_{k=2}^L {\cal L}_{\rm D}^{(k)} \frac{\Delta t}{2n}} 
\right]^n 
  e^{i{\cal L}_{\rm C} \frac{\Delta t}{2}} 
e^{i{\cal L}_{\rm D}^{(1)} \frac{\Delta t}{2}},
\label{eq:16}
\end{gather}
where $n$ is the RESPA integration step parameter. 

\subsection{Parallel Algorithm}

\begin{figure}[htbp]
\centering
\includegraphics[width=0.99\linewidth]{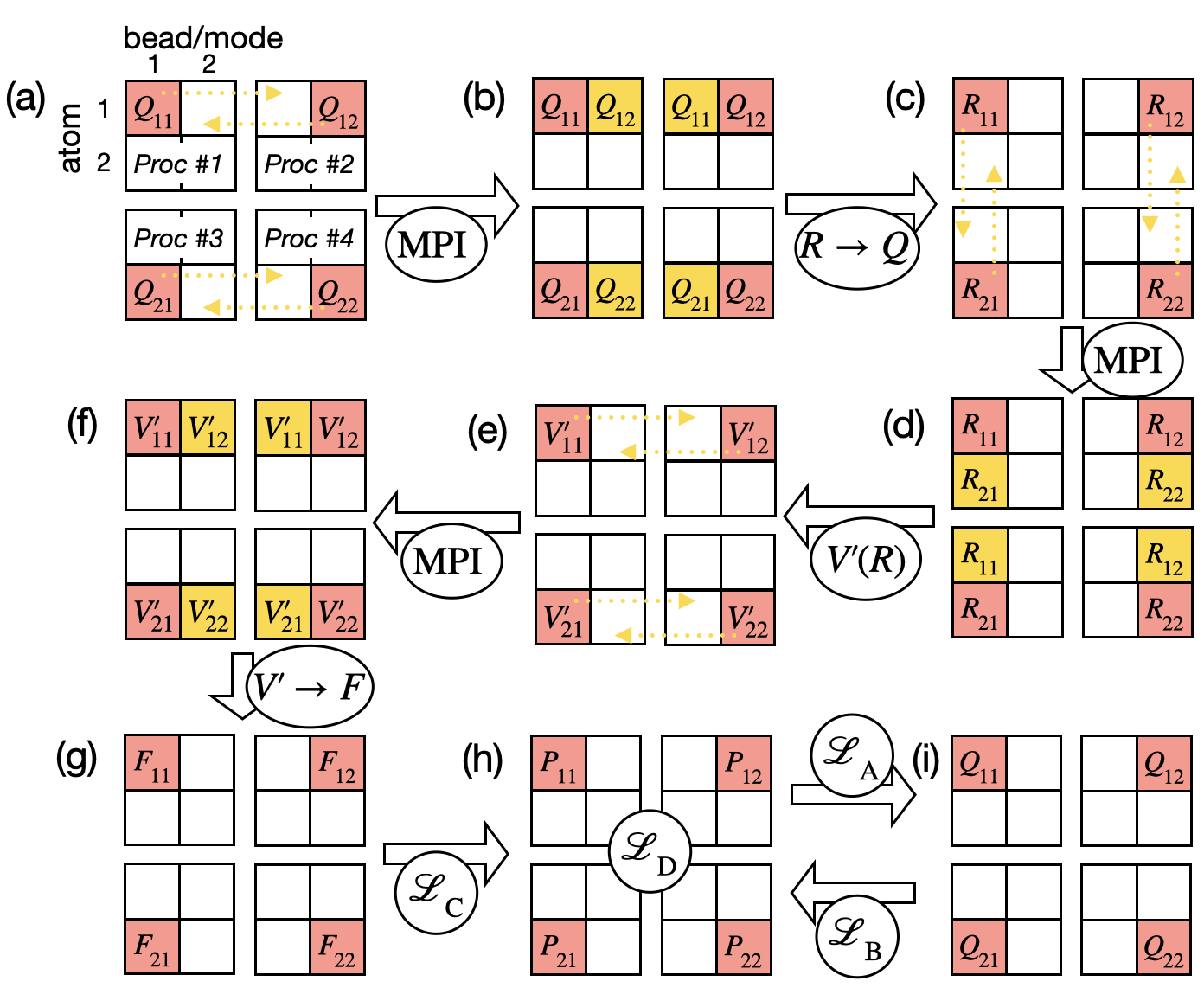}
\caption{The local scheme for the case of a diatomic system with two beads. Orange indicates memory with local values ($Q$, $R$, $V'$, $F$, $P$). Yellow indicates memory written via communication. Yellow arrows show the direction of data transfer, and MPI denotes the communication. $R \rightarrow Q$ and $F \rightarrow V'$ represent normal mode transformations of coordinates and forces, respectively. ${\cal L}_{\rm A}^{}$, ${\cal L}_{\rm B}^{}$, ${\cal L}_{\rm C}^{}$, and ${\cal L}_{\rm D}^{}$ denote the application of the corresponding Liouville operators.}
\label{fig1}
\end{figure}

Figure \ref{fig1} illustrates our new PIMD algorithm using the local scheme, where information pertaining to normal modes and atomic degrees of freedom is uniformly distributed across all available processing units.
For illustrative purposes, the figure shows the simple case of a diatomic system with two beads, parallelized over four processes.

%%%

Each PIMD cycle progresses through the sequence of steps labeled (a) to (i), as shown in Fig.~\ref{fig1}.
The four large squares in each subfigure represent the memory layout of the four respective processing units.
Each unit stores a block of the normal mode coordinates ${\bf Q}$ and momenta ${\bf P}$ corresponding to its assigned modes and atoms. The same applies to the thermostat variables: the coordinates $\boldsymbol{\eta}$ and momenta ${\bf p}$ of the massive NHC thermostats, each coupled to a particular mode of a specific atom.
Inter-process communication is handled via MPI as needed throughout the simulation.

%%%

The operations carried out within a single PIMD cycle are summarized as:
\begin{itemize}
    \item \textbf{Step (a)}: Initialization of the PIMD cycle.
    \item \textbf{Step (b)}: For each atom, the ${\bf Q}$ values are communicated among processes with respect to normal modes.
    \item \textbf{Step (c)}: The normal mode coordinates ${\bf Q}$ are transformed into Cartesian coordinates ${\bf R}$ using Eq.~(\ref{eq:02}).
    \item \textbf{Step (d)}: For each bead, the Cartesian coordinates ${\bf R}$ are communicated among processes with respect to atoms.
    \item \textbf{Step (e)}: The interatomic force routine is invoked to compute the energy gradients $V^\prime({\bf R})$ for each bead.
    \item \textbf{Step (f)}: The computed values $V^\prime({\bf R})$ are shared among the beads through interprocess communication.
    \item \textbf{Step (g)}: The forces $V^\prime({\bf R})$ are transformed into normal mode forces ${\bf F}$ using Eq.~(\ref{eq:15}).
    \item \textbf{Steps (h) and (i)}: The matrices ${\bf Q}$, ${\bf P}$, $\boldsymbol{\eta}$, and ${\bf p}$ are updated according to the equations of motion:
    \begin{itemize}
        \item ${\cal L}_{\rm A}$ updates ${\bf Q}$ using ${\bf P}$ [Eq.~(\ref{eq:11})];
        \item ${\cal L}_{\rm B}$ updates ${\bf P}$ using ${\bf Q}$ [Eq.~(\ref{eq:12})];
        \item ${\cal L}_{\rm C}$ updates ${\bf P}$ using the force ${\bf F}$ [Eq.~(\ref{eq:13})];
        \item ${\cal L}_{\rm D}$ updates $\boldsymbol{\eta}$, ${\bf p}$, and ${\bf P}$ using ${\bf P}$ [Eq.~(\ref{eq:14})].
    \end{itemize}
    \item \textbf{Cycle restart}: Upon completion of these updates, the algorithm returns to Step~(a) to begin the next PIMD cycle.
\end{itemize}
In this way, all updates are performed independently and locally within each process, confined to its corresponding matrix blocks.

\subsection{Applicability to other path integral methods}

Apart from the direct implementation for PIMD described in the Sec. II B, the local scheme is also applicable to approximate quantum dynamics methods such as centroid molecular dynamics (CMD), \cite{cao1994formulation} ring polymer molecular dynamics (RPMD),\cite{craig2004quantum} and Brownian chain molecular dynamics (BCMD).\cite{shiga2018path} These methods can be regarded as extensions or limiting cases of PIMD, with their equations of motion representing simplified forms of those used in PIMD.
Moreover, it is also applicable to path integral hybrid Monte Carlo (PIHMC). \cite{tuckerman1993efficient} It has recently been proposed \cite{thomsen2024self} that PIHMC can be combined with the self-learning technique \cite{nagai2020self} that allows for enhanced sampling and efficient collection of training datasets.

The specific differences from PIMD for each method are as follows:
\begin{itemize}
\item \textbf{CMD}: The thermostats connected to the centroid modes are removed. The centroid masses are set to $\mu_I^{(1)} = M_I^{}$. The masses of the non-centroid modes and their associated NHC thermostats are reduced to small values, promoting rapid equilibration of non-centroid degrees of freedom.
This makes the the centroid dynamics adiabatically separated from the fictitious non-centroid dynamics.
\item \textbf{RPMD}: All thermostats are removed, and the masses of the centroid and non-centroid modes are set equal to the physical atomic masses, $\mu_I^{(k)} = M_I^{}$, for all $I$. This configuration leads to a classical-like dynamics of the ring polymer beads.
\item \textbf{BCMD}: All thermostats are removed. The centroid masses are set as $\mu_I^{(1)} = M_I^{}$, and the masses of the non-centroid modes are set to $\mu_I^{(k)} = \frac{M_I^{}P{\lambda}_k^{}(\Delta t)}{2\beta\hbar}$ for all $I$ with $k \neq 1$, where $\Delta t$ is the time step size. Additionally, the velocities of the non-centroid modes are randomized at every time step according to the Maxwell–Boltzmann distribution, while the centroid-mode velocities remain unchanged. This results in a stochastic dynamics of the ring polymer beads.
\item \textbf{PIHMC}: All thermostats are removed. At specified intervals, the velocities of all modes are randomized according to the Maxwell–Boltzmann distribution. A Metropolis criterion based on the PIMD Hamiltonian is applied to determine the acceptance of the move at the final interval. Due to the detailed balance condition, this samples the correct quantum distribution.
\end{itemize}

\subsection{Heat capacity estimator}

The internal energy is computed as the ensemble average over a PIMD simulation using either the primitive (T) or centroid virial (CV) estimators,
\begin{equation}
U = \left\langle \epsilon_{\rm T} \right\rangle = \left\langle \epsilon_{\rm CV} \right\rangle.
\label{eq:17}
\end{equation}
The primitive estimator is expressed as
\begin{gather}
\epsilon_{\rm T} = \frac{3NP}{2\beta} \nonumber \\ - \frac{P}{2\beta^2\hbar^2} \sum_{\alpha=x,y,z} \sum_{I=1}^N M_I^{} \sum_{s=1}^P \left( R_{I,\alpha}^{(s)} - R_{I,\alpha}^{(s+1)} \right)^2 + \overline{V},
\label{eq:18}
\end{gather}
whereas the CV estimator takes the form
\begin{equation}
\epsilon_{\rm CV} = \frac{3N}{2\beta} + \frac{1}{2} \sum_{\alpha=x,y,z} \sum_{I=1}^N \sum_{s=1}^P \left( R_{I,\alpha}^{(s)} - Q_{I,\alpha}^{(1)} \right) \frac{\partial \overline{V}}{\partial R_{I,\alpha}^{(s)}} + \overline{V},
\label{eq:19}
\end{equation}
where $M_I^{}$ is the mass of atom $I$, and $\overline{V}$ is the bead-averaged potential energy,
\begin{equation}
\overline{V} = \frac{1}{P} \sum_{s=1}^P V\left( \{ {\bf R} \}^{(s)} \right).
\label{eq:20}
\end{equation}

%%%

Similarly, the constant-volume heat capacity can be evaluated using the CV estimator as
\begin{equation}
C_V^{\rm CV} = k_{\rm B} \left[ \beta^2 \left( \left\langle \epsilon_{\rm T} \epsilon_{\rm CV} \right\rangle - \left\langle \epsilon_{\rm T} \right\rangle \left\langle \epsilon_{\rm CV} \right\rangle \right) + \frac{3N}{2} \right],
\label{eq:21}
\end{equation}
or using the double centroid virial (DCV) estimator, which is defined as the sum of two components,
\begin{equation}
C_V^{\rm DCV} = C_V^{\rm DCV1} + C_V^{\rm DCV2},
\label{eq:22}
\end{equation}
with
\begin{gather}
C_V^{\rm DCV1} = k_{\rm B} \left[ \beta^2 \left( \left\langle \epsilon_{\rm CV}^2 \right\rangle - \left\langle \epsilon_{\rm CV} \right\rangle^2 \right) + \frac{3N}{2} \right. \nonumber \\ \left.
- \frac{3\beta}{4} \sum_{\alpha} \sum_{I=1}^N \sum_{s=1}^P \left\langle \Delta R_{I,\alpha}^{(s)} \cdot \frac{\partial \overline{V}}{\partial R_{I,\alpha}^{(s)}} \right\rangle \right],
\label{eq:23}
\end{gather}
and
\begin{gather}
C_V^{\rm DCV2} = - \frac{k_{\rm B} \beta}{4}
\nonumber \\ \times
\sum_{\alpha,\alpha'} \sum_{I=1}^N \sum_{J=1}^N
\sum_{s=1}^P \left\langle \Delta R_{I,\alpha}^{(s)} \frac{\partial^2 \overline{V}}{\partial R_{I,\alpha}^{(s)} \partial R_{J,\alpha'}^{(s)}} \Delta R_{J,\alpha'}^{(s)} \right\rangle,
\label{eq:24}
\end{gather}
where the deviation of the beads from the centroid is given by
\begin{equation}
\Delta R_{I,\alpha}^{(s)} = R_{I,\alpha}^{(s)} - Q_{I,\alpha}^{(1)}.
\label{eq:25}
\end{equation}

%%%

The expressions for $C_V^{\rm CV}$ and $C_V^{\rm DCV}$ were originally derived by Glaesemann and Fried,\cite{glaesemann2002improved} and later extended to multi-element systems by Shiga and Shinoda.\cite{shiga2005calculation}
The second term, $C_V^{\rm DCV2}$, is computationally demanding due to the second derivative of the potential energy, \textit{i.e.}, the Hessian matrix.
To avoid this, numerical differentiation techniques can be employed. This approach was initially proposed by Predescu \textit{et al.},\cite{predescu2003heat} and was subsequently applied to CV- and DCV-based estimators by Yamamoto,\cite{yamamoto2005path} who suggested evaluating $C_V^{\rm DCV2}$ using finite differences of the potential energy.
Recently, Marienhagen and Meier generalized the method to the constant-pressure ensemble.\cite{marienhagen2024calculation}
Another approach to compute the heat capacity of solids has been suggested by Moustafa and Schultz.\cite{moustafa2024generalized}

%%%

In our implementation, we adopt a slightly modified approach, in which the numerical derivative is taken with respect to the atomic forces ${\bf f}$, rather than directly using the potential. This reduces both finite-difference and round-off errors, as it involves a lower-order derivative.
Using a finite-difference approximation to the matrix-vector product ${\bf A}'({\bf x}){\bf y}$ via
\begin{equation}
{\bf A}'({\bf x}){\bf y} \approx \pm \frac{1}{\delta} \left[ {\bf A}({\bf x} \pm \delta{\bf y}) - {\bf A}({\bf x}) \right],
\end{equation}
we numerically approximate $C_V^{\rm DCV2}$ as
\begin{gather}
C_V^{\rm DCV2} = \frac{k_{\rm B} \beta}{8P} \sum_{\alpha} \sum_{I=1}^N \sum_{s=1}^P \left\langle \frac{\Delta R_{I,\alpha}^{(s)}}{\delta^{(s)}} 
\right. \nonumber \\ \left.
\left\{ f_{I,\alpha}^{(s)}\left( \{{\bf R} + \delta \Delta{\bf R} \}^{(s)} \right) - f_{I,\alpha}^{(s)}\left( \{{\bf R} - \delta \Delta{\bf R} \}^{(s)} \right) \right\} \right\rangle,
\label{eq:26}
\end{gather}
where the force at the displaced positions is defined as
\begin{gather}
f_{I,\alpha}^{(s)} \left( \{ {\bf R} \pm \delta \Delta {\bf R} \}^{(s)} \right)
\nonumber \\
= -\frac{\partial V \left( R_{1,x}^{(s)} \pm \delta^{(s)} \Delta R_{1,x}^{(s)}, \cdots,
R_{N,z}^{(s)} \pm \delta^{(s)} \Delta R_{N,z}^{(s)}
\right)}{\partial \left( R_{I,\alpha}^{(s)} \pm \delta^{(s)} \Delta R_{I,\alpha}^{(s)} \right)}.
\label{eq:27}
\end{gather}

%%%

%
The DCV estimator requires two additional force evaluations per sample, which are not part of the standard PIMD force computation.
However, $C_V^{\rm DCV2}$ tends to converge with much fewer samples than $C_V^{\rm DCV1}$ in all cases studied.
Thus, $C_V^{\rm DCV1}$ is computed at every time step, while $C_V^{\rm DCV2}$ is computed only once every 100 steps. In this way, the overall number of force evaluations was increased just by 2\%.

%%%

The displacement parameter $\delta^{(s)}$ must be chosen to balance numerical stability and accuracy. To prevent large shifts in regions of small displacement, while also keeping the components $\delta \Delta {\bf R}$ small (less than a threshold $d$), we define
\begin{equation}
\delta^{(s)} = \frac{d}{\max\left( |\Delta R_{1,x}^{(s)}|, \cdots, |\Delta R_{N,z}^{(s)}|, d \right)},
\label{eq:28}
\end{equation}
with $d = 10^{-4}$ bohr. This choice yields numerically stable results, and the accuracy is not particularly sensitive to the specific value of $d$.

{
We note that conventional  classical estimators are obtained in the $P=1$ case.
From Equations (\ref{eq:18}) and (\ref{eq:19})
the classical energy estimator corresponds to
\begin{equation}
\epsilon^{\rm cl} = \frac{3N}{2\beta} + V,
\end{equation}
and from Equations (\ref{eq:21}) and (\ref{eq:22}) the classical heat capacity corresponds to
\begin{equation}
C_V^{\rm cl} = k_{\rm B} \left[ \beta^2 \left( \left\langle V^2 \right\rangle - \left\langle V\right\rangle^2 \right) + \frac{3N}{2} \right].
\end{equation}
The latter was used for classical MD.
}

\section{Computational details}

\subsection{High-Dimensional Neural Network Potentials}

{
The selection of training data is crucial in machine learning techniques.\cite{sadus2023molecular} For the sake of the construction of HDNNPs, we employed in part the self-learning PIHMC method, \cite{nagai2020self,kobayashi2021self, thomsen2024self} which ensures that the DFT configurations used for training are sampled from physically meaningful ensembles. This approach increases the reliability of the HDNNP, and thus the consistency of thermodynamic averages obtained from the HDNNP-based PIMD simulations.
}

The second-generation HDNNP, NN-RPBE-D3, was adopted from previous studies on the water/ZnO interface.\cite{quaranta2017, quaranta2018, hellstrom2019, quaranta2019} It was trained on DFT data using the RPBE exchange-correlation functional,\cite{hammer_improved_1999} combined with the D3 dispersion corrections.\cite{grimme_consistent_2010, grimme_effect_2011} The symmetry function parameters for hydrogen and oxygen atoms are listed in Table \ref{tab1}. The root mean square error (RMSE) of the final fit is 1.0 meV/atom for energies and 74.3 meV/bohr for forces, as evaluated using the \texttt{RuNNer} software. \cite{runner2018}

\begin{table}[htbp]
\caption{Parameters of the radial and angular symmetry functions \cite{quaranta2017} for the HDNNPs of water, in atomic units (a.u.). 
For all functions the radial cutoff distance was set to $R_{\rm cutoff}^{}=12.00$ a.u.}
\begin{ruledtabular}
\begin{tabular}{cccccc}
Atom 1 & Atom 2 & \multicolumn{2}{c}{$R_{\rm shift}^{}$} & \multicolumn{2}{c}{$\eta$} \\
\hline
H & H & \multicolumn{2}{c}{0.0} & \multicolumn{2}{c}{$\{0.0009,0.01,0.02,0.05,0.1\}$} \\
H & H & \multicolumn{2}{c}{1.5} & \multicolumn{2}{c}{$\{0.05,0.1,0.2,0.5,0.9 \}$}\\
H & O & \multicolumn{2}{c}{0.0} & \multicolumn{2}{c}{$\{0.0009,0.01,0.02,0.05,0.1\}$} \\
H & O & \multicolumn{2}{c}{0.8} & \multicolumn{2}{c}{$\{0.05,0.1,0.2,0.5,0.9\}$} \\
O & H & \multicolumn{2}{c}{0.0} & \multicolumn{2}{c}{$\{0.0009,0.01,0.02,0.05,0.1\}$} \\
O & H & \multicolumn{2}{c}{0.8} & \multicolumn{2}{c}{$\{0.05,0.1,0.2,0.5,0.9\}$} \\
O & O & \multicolumn{2}{c}{0.0} & \multicolumn{2}{c}{$\{0.0009,0.01,0.02,0.05,0.1\}$} \\
O & O & \multicolumn{2}{c}{3.0} & \multicolumn{2}{c}{$\{0.05,0.1,0.2,0.5,0.9\}$} \\
\hline\hline
Atom 1 & Atom 2 & Atom 3 & $\eta$ & $\lambda$ & $\zeta$ \\
\hline
O & H & H & 0.07 & $\{-1.0,1.0\}$ & 1.0 \\
O & H & H & 0.03 & $\{-1.0,1.0\}$ & 1.0 \\
O & H & H & 0.01 & $\{-1.0,1.0\}$ & 4.0 \\

H & O & H & 0.07 & 1.0 & 1.0 \\
H & O & H & 0.03 & $\{-1.0,1.0\}$ & 1.0 \\
{H} & O & H & 0.01 & $\{-1.0,1.0\}$ & 4.0 \\
{H} & O & O &
{0.03} & $-1.0$ & 1.0 \\
H & O & O & 0.001 & $\{-1.0,1.0\}$ & 4.0 \\
O & O & H & 0.001 & $\{-1.0,1.0\}$ & 4.0 \\
{O} & O & H & 0.03 & $\{-1.0,1.0\}$ & 
{1.0} \\
O & O & O & 0.001 & 1.0 & 4.0 \\
\end{tabular}
\end{ruledtabular}
\label{tab1}
\end{table}

The NN-revPBE0-D3 potential was constructed in the present work using the same symmetry function parameters as the NN-RPBE-D3 potential.
In this case, the training set was collected from DFT calculations using the \texttt{cp2k} software \cite{kuhne_cp2k:_2020} using the revPBE0 exchange-correlation functional \cite{perdew1996generalized,PhysRevLett.80.890,adamo1999toward} with D3 dispersion corrections.\cite{grimme_consistent_2010, grimme_effect_2011}
For this purpose, a series of self-learning PIHMC simulations \cite{thomsen2024self} were performed in the NPT ensemble for liquid and supercritical water under the following conditions:
(298 K, 1 bar), (358 K, 1 bar), (423 K, 100 bar), (573 K, 100 bar), (673 K, 500 bar), and (753 K, 2540 bar).
In addition, a MD simulation was performed in the NVT ensemble at 3000 K with a density of $\rho = 1$ g/cm$^3$.
From each of these seven simulations, 320 structures were extracted and included in the training and test sets.
The HDNNP was optimized for 200 epochs using the \texttt{n2p2} software.\cite{n2p2220}
The RMSE in the test set is 0.83 meV/atom for energies and 44.0 meV/bohr for force components.

\begin{figure}[htbp]
\centering
\includegraphics[width=0.99\linewidth]{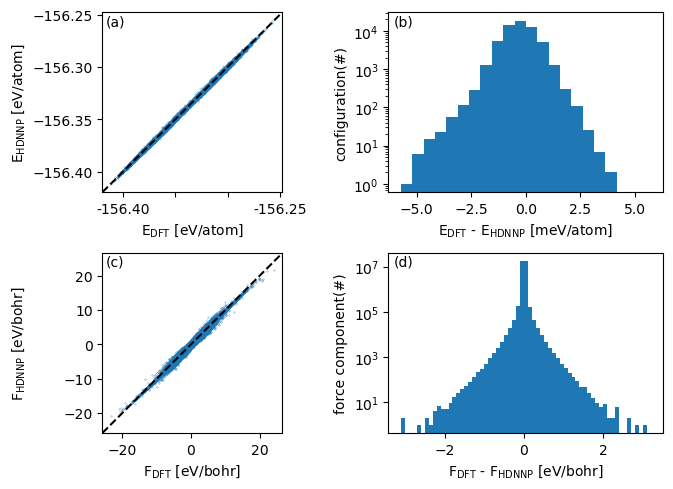}
\caption{
Error analysis of the NN-revPBE0-D3 model using 59,520 water structures, which are not included in the training or test sets. 
(a) Comparison between DFT and HDNNP energies per atom, the black dashed line indicates E$_{\mathrm{DFT}}$=E$_{\mathrm{HDNNP}}$.
(b) The distribution of the difference of the DFT and HDNNP energies per atom.
(c) Same as (a) for the force components from DFT and HDNNP. (d) Same as (b) for the difference of the force components.
}
\label{fig2}
\end{figure}

For further assessment of the NN-revPBE0-D3 model, we took 9,920 water structures each (59,520 structures in total) from the simulations at 298 K, 358 K, 423 K, 573 K, 673 K and 753 K to evaluate
the RMSE of the energies,
\begin{equation}
\sigma^{\mathrm{at}}_{\mathrm{E}}=\sqrt{\frac{1}{N}\sum_{i=1}^{N}
\left(\mathrm{E}_{\mathrm{DFT}}^{i}-\mathrm{E}_{\mathrm{HDNNP}}^{i}\right)^2}=0.75 \frac{\mathrm{meV}}{\mathrm{atom}},
\end{equation}
and the RMSE of the force components,
\begin{equation}
\sigma_{\mathrm{F}}^{\mathrm{at}}=\sqrt{\frac{1}{3N}\sum_{i=1}^{N}\sum^{x,y,z}_{\alpha}\left(\mathbf{F}_{\mathrm{DFT}}^{i,\alpha}-\mathbf{F}_{\mathrm{HDNNP}}^{i,\alpha}\right)^2}=40.1 \frac{\mathrm{meV}}{\mathrm{bohr}},
\end{equation}
which are very similar to the respective RMSE values of the test set.
Furthermore, these RMSE values are in the same order of magnitude as those reported for the NN-RPBE-D3 model.
In Figure \ref{fig2}(a) and (c) we have plotted the correlations between the NN-revPBE0-D3 energies and force components and the reference revPBE0-D3 DFT energies and force components, respectively.
The distribution of the energy and force differences is plotted in Figures \ref{fig2}(b) and (d), respectively.
It can be seen that the NN-revPBE0-D3 model agrees very well with the reference revPBE0-D3 DFT data, both for energies and forces, without having any significant outliers.

\subsection{PIMD Setup}

The PIMD simulations used to compute $C_V^{}$ in the NVT ensemble were performed using the NN-RPBE-D3 and NN-revPBE0-D3 potentials for a system of 256 water molecules represented with 128 beads, enclosed in a periodic cubic box at a density of 1.0 g/cm$^3$.  
This number of beads (128) was chosen based on previous studies using the SPC/F2 force field, \cite{lobaugh1997quantum} where convergence of $C_V^{}$ was achieved.\cite{shiga2005calculation}
The temperature was maintained at 300 K using massive NHC thermostats, and the time step was set to 0.125 fs.  
Classical MD simulations were performed under identical conditions but with a single bead.
For benchmarking parallel performance, additional short PIMD simulations were performed using either 32 or 128 beads for systems containing 256 and 2048 water molecules.  

Thermostat masses were set as follows: 
\( m_1 = \tau_{\rm sys}^2 / \beta \), 
where \( \tau_{\rm sys} \) is a parameter representing the vibrational timescale of the system (10~fs); 
and \( m_j = \beta \hbar^2 / P \) for \( 2 \le j \le L \), with the Nosé--Hoover chain (NHC) length set to \( L = 4 \). 
A RESPA integration step parameter of \( n = 5 \) was used.

All simulations were performed using the \texttt{PIMD} software, \cite{shiga2025pimd} which features parallel implementations in both local and shared schemes and includes interfaces to the \texttt{n2p2} software \cite{n2p2220} and the \texttt{cp2k} software. \cite{kuhne_cp2k:_2020}

\section{Results and Discussion}

\subsection{Parallel performance}

\begin{figure}[htbp]
\centering
\includegraphics[width=0.99\linewidth]{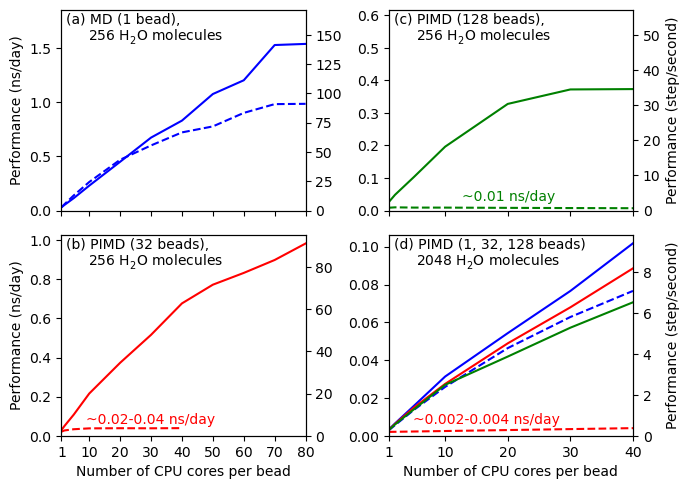}
\caption{Parallel performance compared between the local (solid lines) and shared (dashed lines) schemes for liquid water systems, with a PIMD time step of 0.125 fs.
Computational speed (nanoseconds per elapsed day and steps per elapsed second) is shown against the number of CPU cores used per bead.
Results are shown for a 256-molecule model: (a) single-bead MD (blue), (b) 32-bead PIMD (green), and (c) 128-bead PIMD (red).
Panel (d) shows results for a 2048-molecule model, using the same color scheme as (a)--(c).}
\label{fig3}
\end{figure}

Figure \ref{fig3} compares the parallel performance of both, the shared and the local scheme, in the PIMD simulations. The evaluation was conducted on JAEA's \texttt{HPE SGI8600} system, in which each compute node is equipped with two \texttt{Intel Xeon Gold 6242R} processors (3.1 GHz, 20 CPU cores each). For water simulations containing 256 or 2048 molecules, the computational speed was measured in nanoseconds per elapsed day (ns/day) and the number of steps per elapsed second (step/sec).

For a comparison between classical MD (1 bead) and PIMD (32 and 128 beads), the horizontal axis of the figure is defined as the number of CPU cores used divided by the number of beads. The maximum of 5,120 CPU cores corresponds to the configuration with 128 beads and 40 CPU cores per bead.

In the case of classical MD, there is little difference in performance between the shared and local schemes as long as the number of CPU cores used is small. In this regime, the total computation time per step is relatively long, and the primary bottleneck is the evaluation of the interatomic energy and forces. Since this part of the calculation in the local scheme has not been significantly modified from the shared one, parallelization does not lead to noticeable performance differences.

As the number of CPU cores increases, a clear performance gap emerges between the local and shared schemes. Specifically, for the 256-molecule model, the performance of the shared scheme begins to degrade beyond 20 CPU cores, and for the 2048-molecule model, this occurs beyond 10 CPU cores. In contrast, the local scheme improves the linear scaling up to a larger number of CPU cores. This difference arises from the fact that, in the local scheme, the time evolution of positions and momenta for both atoms and thermostats is parallelized on a per-atom and per-mode basis, and this parallelization strategy works effectively.

When the number of CPU cores becomes very large, even the local scheme eventually loses linear scalability, and performance starts to decline. For the 256-molecule model, scalability saturates at around 75 CPU cores, corresponding to a workload of approximately 10 atoms per CPU core. In this region, the overhead from MPI communication, which is required to update all atoms on each CPU core, becomes dominant, outweighing the benefits of further parallel decomposition.

The observed non-uniform scaling behavior arises from an imbalance in the computational workload distributed across CPU cores. Although the local scheme assigns tasks on a per-atom basis, the HDNNP employed in this study uses a different number of symmetry functions for each atom type, such as 31 for oxygen and 27 for hydrogen, resulting in an inherent load imbalance that may degrade overall performance.

Unlike in MD simulations, the local scheme performs significantly better than the old one in PIMD simulations.
Moreover, as in the case of MD, parallel efficiency in PIMD simulations tends to improve with increasing system size.
This is evidenced by the observation that the parallel efficiency curve becomes more linear as the number of atoms increases.
In particular, the 2048-molecule model shows substantially better linearity compared to the 256-molecule model.
These results demonstrate that the local scheme is highly effective for PIMD, primarily because it enables efficient mode-level parallelization.

In terms of parallel efficiency with respect to the number of CPU cores (\textit{i.e.}, the ratio of performance between single-core and multi-core calculations, respectively, for MD and PIMD), PIMD exhibits higher efficiency than MD. This is because PIMD involves a larger computational workload, which can be more effectively distributed across multiple CPU cores.

However, when comparing the parallel efficiency per bead (defined as the parallel efficiency multiplied by the performance ratio between the $n$-core MD and $nP$-core PIMD calculations), PIMD shows lower efficiency than MD.
For example, when 40 CPU cores per bead are used, the parallel efficiency for the 256- and 2048-molecule MD models is 68.3\% and 67.1\%, respectively. On the other hand, for PIMD, the efficiency becomes 58.8\% and 77.3\%, respectively, with 32 beads, and 35.5\% and 70.9\%, respectively, with 128 beads.
This reduction in efficiency is mainly due to the overhead from MPI communication in the local scheme. As the number of atoms decreases and the number of beads increases, the proportion of time spent on communication relative to computation tends to increase, leading to overall performance degradation.

\subsection{{Functional dependence}}

\begin{figure*}[htbp]
\centering    \includegraphics[width=0.99\textwidth]{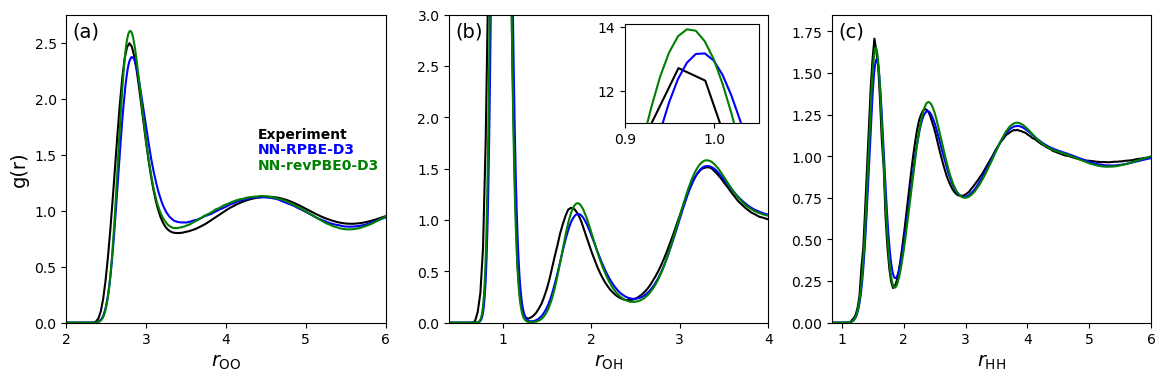}
\caption{{(a) O-O, (b) O-H, and (c) H-H RDFs for the PIMD calculations carried out in this study. In (b) the insert shows the location of the first O-H RDF peak, due to its difference in height from the other two peaks. The results in blue and green are obtained with the NN-RPBE0-D3 and NN-revPBE0-D3 potentials, respectively. The calculated RDFs are compared with the experimental results in black from Ref. \onlinecite{soper_radial_2013} for O-O and Ref. \onlinecite{soper_radial_2000} for O-H and H-H. A comparison with the results from NN-MD simulations and results from DFT-based PIMD and PIHMC-MIX are given in Figure S1 in the Supporting Material.}}
\label{fig7}
\end{figure*}

{
Because the heat capacity of water is strongly influenced by structural fluctuations, it is essential to use a DFT functional that can accurately reproduce the structure of liquid water. Our previous studies \cite{machida2018nuclear,thomsen2024self} systematically examined the performance of several DFT functionals --- including BLYP-D2, SCAN, rev-vdW-DF2, and optB88-vdW --- in path integral simulations. Among these, RPBE-D3 was found to provide the best agreement with experimental radial distribution functions (RDFs), which motivated its selection for the present heat capacity calculations.
}

{
In this study, we also employed revPBE0-D3, a hybrid functional that incorporates a fraction of Hartree–Fock exchange. Hybrid functionals are generally expected to improve the description of electronic exchange interactions, and, as shown below, this choice resulted in better agreement with experimental heat capacity data, demonstrating the advantage of hybrid functionals for modeling water thermodynamics.
}

{
To illustrate the structural accuracy of the two functionals used, we have added Figure 4(a)–(c), which compares the RDFs of O–O, O–H, and H–H atom pairs obtained from NN-RPBE-D3 and NN-revPBE0-D3 PIMD simulations with experimental data.\cite{soper_radial_2000,soper_radial_2013} Both functionals show good agreement with experiment, with revPBE0-D3 offering a slightly better match.
}

{
Furthermore, panels (a)–(f) of Figure S1 in the Supporting Material demonstrate that PIMD simulations yield RDFs that agree more closely with experimental results than those from classical MD simulations, highlighting the importance of including nuclear quantum effects.
}

\subsection{Heat capacity}

\begin{figure}[htbp] % [htbp]
\centering
\includegraphics[width=0.99\linewidth]{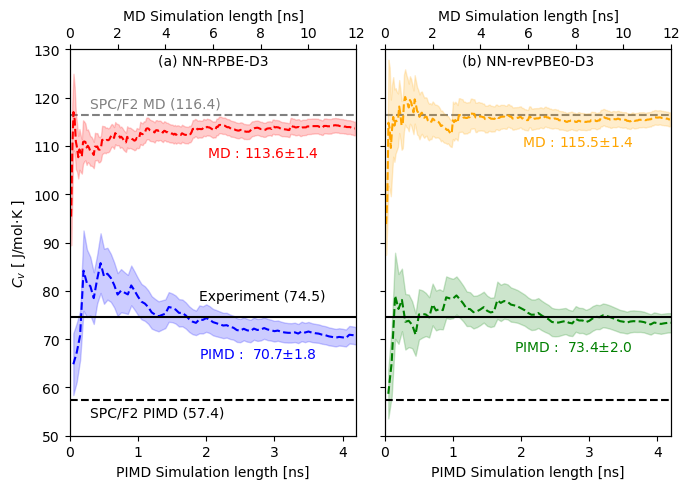}
\caption{Cumulative average of $C_V^{}$ calculated using the DCV estimator [Eq.(\ref{eq:22})] from 12 ns MD and 4.2 ns PIMD trajectories.
The left panel shows results obtained with the NN-RPBE-D3 potential: MD (dashed red lines) and PIMD (solid blue lines).
The right panel shows results using the NN-revPBE0-D3 potential: MD (dashed orange lines) and PIMD (solid green lines).
Shaded regions represent statistical uncertainties estimated using the method of Flyvbjerg and Petersen. \cite{flyvbjerg1989error}
For comparison, converged values from classical MD (dashed grey lines) and PIMD (solid grey lines) simulations using the SPC/F2 force field \cite{shiga2005calculation} are shown.
The experimental value is shown as a solid black line.}
\label{fig4}
\end{figure}

\begin{figure}[htbp]
\centering    \includegraphics[width=0.99\linewidth]{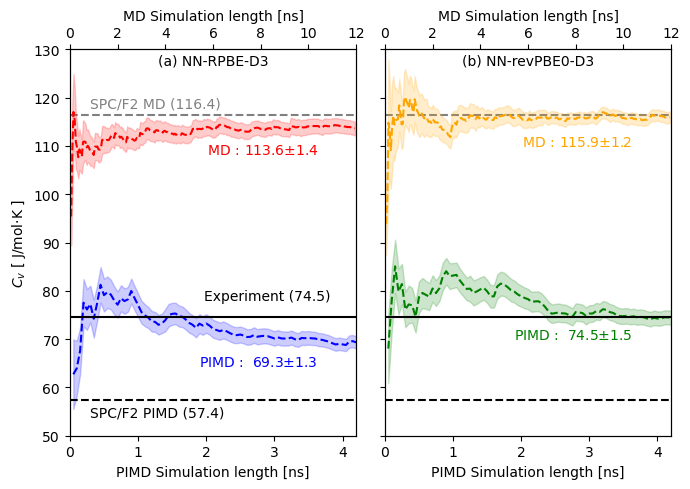}
\caption{Cumulative average of $C_V^{}$ calculated using the CV estimator [Eq.(\ref{eq:21})] from 12 ns MD and 4.2 ns PIMD trajectories. Otherwise the same as Figure \ref{fig4}.}
\label{fig5}
\end{figure}

The cumulative averages of the heat capacities $C_V^{\rm CV}$ and $C_V^{\rm DCV}$ computed using each method are shown in Figures \ref{fig4} and \ref{fig5}, respectively.
Note that $C_V^{\rm CV}$ and $C_V^{\rm DCV}$ are the same for MD, since Eqs.~(\ref{eq:21}) and (\ref{eq:22}) are identical with $P=1$.
Table \ref{tab2} summarizes the final average values along with their statistical errors and compares them with the IAPWS-95\cite{IAPWS1995} value calculated at $T=298.15$ K and $\rho=0.997$ g/ml.
Although $C_V^{\rm CV}$ generally exhibits slightly smaller statistical fluctuations than $C_V^{\rm DCV}$, both methods ultimately converge to consistent values.

\begin{table}[htbp]
\caption{Heat capacity of liquid water in J/mol K.}
\begin{center}
\begin{tabular}{cccc}
\hline\hline
 \hspace{0mm} method \hspace{0mm} &
 \hspace{0mm} NN-RPBE-D3 \hspace{0mm} & \hspace{0mm} NN-revPBE0-D3 \hspace{0mm} & \\
\hline
 MD                   & $113.6\pm 1.4$ & $115.5\pm 1.4$ & \\
 PIMD (CV) & $69.3\pm 1.3$ & $74.5\pm 1.5$ & \\
 PIMD (DCV)  & $70.7\pm 1.8$ & $73.4\pm 2.0$ & \\
 \hline
 Exptl [\onlinecite{IAPWS1995}]  &  & & \hspace{0mm} $74.545^a$  \hspace{0mm} \\
\hline\hline
\multicolumn{4}{l}{$^a$The IAPWS-95 value at $T=298.15$ K, $\rho=0.997$ g/ml.}
\end{tabular}
\end{center}
\label{tab2}
\end{table}

The heat capacities obtained from PIMD simulations using NN-RPBE-D3 and NN-revPBE0-D3 are in very good agreement with the experimental data.
In particular, the results based on NN-revPBE0-D3 fall within the statistical error of the experimental value.
As discussed in the Introduction, accurate estimation of the heat capacity requires a correct balance between two opposing contributions: the increase due to the structural flexibility of hydrogen bonds and the decrease caused by nuclear quantum effects.
The close agreement with the experiment indicates that this balance is appropriately captured by PIMD simulations using HDNNPs.
{
Both RPBE-D3 and revPBE0-D3 functionals effectively capture the structural fluctuations associated with the hydrogen-bond network in liquid water, which are critical for accurately predicting its heat capacity. The improved performance of revPBE0-D3 likely stems from the inclusion of a fraction of exact Hartree–Fock exchange, which enhances the description of the electronic structure and strengthens the accuracy of intermolecular interactions.
}

In contrast, classical MD simulations using the same HDNNPs significantly overestimate the heat capacity $C_V^{}$.
This overestimation arises from the neglect of quantum effects, which disrupts the balance described above.
Water molecules possess three intramolecular vibrational degrees of freedom whose energy level spacings ($\hbar\omega$) are significantly larger than the thermal energy ($k_{\rm B}T$), leading to pronounced quantum suppression.
Therefore, an overestimation of approximately $3 k_{\rm B}$ is not unexpected if these modes are treated classically.
However, the actual differences in the heat capacity between PIMD and classical MD for NN-RPBE-D3 and NN-revPBE0-D3 are about 5.3 $k_{\rm B}^{}$ (44 J/mol K) and 4.9 $k_{\rm B}^{}$ (41 J/mol K), respectively, which exceed this expected value.
This indicates that quantum effects play a significant role not only in the intramolecular degrees of freedom, but also in the intermolecular degrees of freedom, particularly in the fluctuations of the hydrogen-bond network.

Previous PIMD studies using classical force fields have shown that the predicted heat capacity of liquid water strongly depends on the specific model employed. \cite{berta2020nuclear}
For example, in non-polarizable flexible models it has been reported that SPC/F2\cite{shiga2005calculation} (57.4 J/mol$\cdot$K) and q-TIP4P/F\cite{ceriotti2011accelerating} (65.7 J/mol$\cdot$K) underestimate while q-SPC/Fw \cite{paesani2006accurate} (87.9 J/mol$\cdot$K) overestimate the heat capacity of liquid water.
The present results using HDNNPs highlight the limitations of classical force fields and demonstrate that more expressive models are essential.
HDNNPs are capable of naturally representing a range of physical effects, including intramolecular bonding, intermolecular interactions, charge transfer, and molecular polarization, all of which are crucial for accurately describing liquid water.

It is worth noting that PIMD simulations with HDNNPs exhibit somewhat less uniform energy fluctuations compared to those using classical force fields, which leads to longer convergence times.
However, this behavior can be interpreted as a reflection of the enhanced fidelity of HDNNPs to reproduce realistic physical fluctuations in the system.

\begin{figure}[htbp]
\centering
\includegraphics[width=0.99\linewidth]{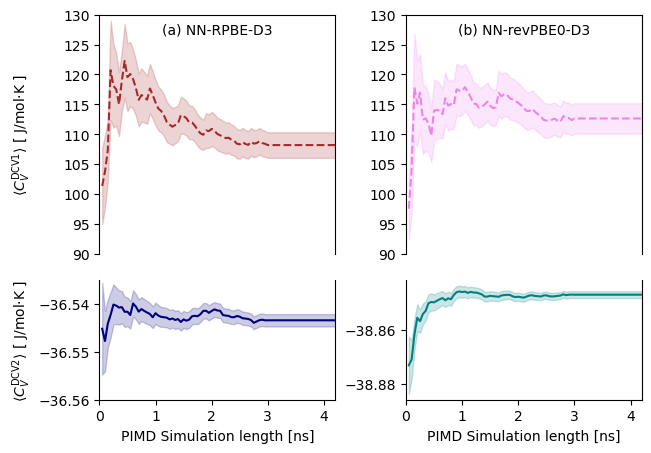}
\caption{Cumulative averages of $C_V^{\rm DCV1}$  and $C_V^{\rm DCV2}$ from 4.2 ns PIMD trajectories. The left panel shows results obtained with the NN-RPBE-D3 potential, $C_V^{\rm DCV1}$ (dashed brown lines) and $C_V^{\rm DCV2}$ (solid navy lines).
The right panel shows results using the NN-revPBE0-D3 potential: $C_V^{\rm DCV1}$ (dashed violet lines) and $C_V^{\rm DCV2}$ (solid teal lines).
Shaded regions represent statistical uncertainties estimated using the method of Flyvbjerg and Petersen. \cite{flyvbjerg1989error}}
\label{fig6}
\end{figure}

Figure \ref{fig6} shows the cumulative average of the heat capacity for the efficient DCV estimator, $C_V^{\rm DCV}$, with the contributions from the first term [$C_V^{\rm DCV1}$, Eq.(\ref{eq:23})] and the second term [$C_V^{\rm DCV2}$, Eq.(\ref{eq:24})] separately illustrated.
It is evident that $C_V^{\rm DCV2}$ converges statistically much more rapidly than $C_V^{\rm DCV1}$.
Therefore, the strategy adopted in this study, \textit{i.e.}, evaluating the $C_V^{\rm DCV1}$ at every time step using force information and computing $C_V^{\rm DCV2}$ every 100 steps by numerical differentiation, is efficient and effective.
Importantly, this approach allows for the inclusion of $C_V^{\rm DCV2}$ without significantly increasing the overall computational cost, making it a practical and favorable method for accurate evaluation of the heat capacity.

\section{Conclusion}

In this study, we have demonstrated that PIMD simulations using HDNNPs trained on first-principles data provide a highly accurate and reliable framework for evaluating the heat capacity of liquid water. By incorporating both the flexibility of the hydrogen-bond network and the quantum nature of atomic nuclei, our approach successfully reproduces experimental values of the heat capacity under ambient conditions.

The new parallel algorithm that uses the local scheme has significantly improved the performance and scalability of PIMD simulations. This has enabled ns-scale simulations required for statistically robust estimates of thermodynamic properties. Performance improvements were particularly pronounced in systems with thousands of water molecules, confirming the suitability for large-scale quantum simulations.

The heat capacity results obtained with HDNNPs based on the RPBE-D3 and revPBE0-D3 functionals show clear advantages over MD and PIMD simulations using CFFs. Classical MD consistently overestimates $C_V^{}$ by failing to account for quantum suppression of intramolecular vibrational modes, while PIMD captures the delicate balance between structural flexibility and quantum effects. The excess contribution of quantum structural fluctuations to the heat capacity beyond what can be attributed solely to three intramolecular vibrational degrees of freedom of water molecules highlights the importance of accurately modeling complex hydrogen-bond dynamics.
These findings underscore the critical role of quantum effects in determining the thermodynamic properties of water in general and affirm the suitability of HDNNP-based PIMD as a powerful tool for predictive simulations. 

The constant-volume heat capacity was the focus of this work.
{We believe that this serves as a benchmark at ambient conditions and lays the groundwork for future studies at different thermodynamic states.}
To obtain the constant-pressure heat capacity, it may be necessary to establish an efficient parallel algorithm for PIMD in the NPT ensemble. Once this challenge is addressed, it is expected that constant-pressure heat capacity can also be accurately computed using HDNNP-based PIMD, as demonstrated in this work.

\begin{acknowledgments}
M.S. and B.T. thank the JSPS Grant-in-Aid for Scientific Research (Grant Numbers 23K04670, 23H01273, 24K01145, 24K01408, and 25K01629) for financial support.  J.E. and J.B. thank the Deutsche Forschungsgemeinschaft (DFG) for funding in CRC 1633 (C04, project number 510228793) and for support under Germany's Excellence Strategy – EXC 2033 RESOLV (project number 390677874). M.S. gratefully acknowledges the support from EXC 2033 RESOLV for the Visiting Professor Program at Ruhr Universit\"at Bochum. The calculations were conducted using the supercomputing facilities of Japan Atomic Energy Agency, with primary use of the \texttt{HPE SGI8600} system.
 \\

\end{acknowledgments}

%%%%%%%%%%%%%%%%%%%%%%%%%%%%%%%%%%%%%%%%%%%%%%%%%%%%%%%%%%%%%%%%%%%%%%%%
\section*{Supplementary Material}
%%%%%%%%%%%%%%%%%%%%%%%%%%%%%%%%%%%%%%%%%%%%%%%%%%%%%%%%%%%%%%%%%%%%%%%%

An extended comparison of the RDFs from MD and PIMD simulations with experimental data is provided in the supplementary material.
\ \\

%%%%%%%%%%%%%%%%%%%%%%%%%%%%%%%%%%%%%%%%%%%%%%%%%%%%%%%%%%%%%%%%%%%%%%%%
\section*{Data Availability}
%%%%%%%%%%%%%%%%%%%%%%%%%%%%%%%%%%%%%%%%%%%%%%%%%%%%%%%%%%%%%%%%%%%%%%%%

The data that support the findings of this study are available within
 the article.

%%%%%%%%%%%%%%%%%%%%%%%%%%%%%%%%%%%%%%%%%%%%%%%%%%%%%%%%%%%%%%%%%%%%%%%%
%\bibliographystyle{ieeetr}
\bibliography{xmpi.bib}

%aipnum4-2.bst 2019-01-14 (MD) hand-edited version of apsrev4-1.bst
\begin{thebibliography}{72}%
\makeatletter
\providecommand \@ifxundefined [1]{%
 \@ifx{#1\undefined}
}%
\providecommand \@ifnum [1]{%
 \ifnum #1\expandafter \@firstoftwo
 \else \expandafter \@secondoftwo
 \fi
}%
\providecommand \@ifx [1]{%
 \ifx #1\expandafter \@firstoftwo
 \else \expandafter \@secondoftwo
 \fi
}%
\providecommand \natexlab [1]{#1}%
\providecommand \enquote  [1]{``#1''}%
\providecommand \bibnamefont  [1]{#1}%
\providecommand \bibfnamefont [1]{#1}%
\providecommand \citenamefont [1]{#1}%
\providecommand \href@noop [0]{\@secondoftwo}%
\providecommand \href [0]{\begingroup \@sanitize@url \@href}%
\providecommand \@href[1]{\@@startlink{#1}\@@href}%
\providecommand \@@href[1]{\endgroup#1\@@endlink}%
\providecommand \@sanitize@url [0]{\catcode `\\12\catcode `\$12\catcode
  `\&12\catcode `\#12\catcode `\^12\catcode `\_12\catcode `\%12\relax}%
\providecommand \@@startlink[1]{}%
\providecommand \@@endlink[0]{}%
\providecommand \url  [0]{\begingroup\@sanitize@url \@url }%
\providecommand \@url [1]{\endgroup\@href {#1}{\urlprefix }}%
\providecommand \urlprefix  [0]{URL }%
\providecommand \Eprint [0]{\href }%
\providecommand \doibase [0]{https://doi.org/}%
\providecommand \selectlanguage [0]{\@gobble}%
\providecommand \bibinfo  [0]{\@secondoftwo}%
\providecommand \bibfield  [0]{\@secondoftwo}%
\providecommand \translation [1]{[#1]}%
\providecommand \BibitemOpen [0]{}%
\providecommand \bibitemStop [0]{}%
\providecommand \bibitemNoStop [0]{.\EOS\space}%
\providecommand \EOS [0]{\spacefactor3000\relax}%
\providecommand \BibitemShut  [1]{\csname bibitem#1\endcsname}%
\let\auto@bib@innerbib\@empty
%</preamble>
\bibitem [{\citenamefont {Eisenberg}\ and\ \citenamefont
  {Kauzmann}(1969)}]{eisenberg1969structure}%
  \BibitemOpen
  \bibfield  {author} {\bibinfo {author} {\bibfnamefont {D.}~\bibnamefont
  {Eisenberg}}\ and\ \bibinfo {author} {\bibfnamefont {W.}~\bibnamefont
  {Kauzmann}},\ }\href@noop {} {\emph {\bibinfo {title} {The structure and
  properties of water}}}\ (\bibinfo  {publisher} {Oxford University Press},\
  \bibinfo {year} {1969})\BibitemShut {NoStop}%
\bibitem [{\citenamefont {Bukowski}\ \emph {et~al.}(2007)\citenamefont
  {Bukowski}, \citenamefont {Szalewicz}, \citenamefont {Groenenboom},\ and\
  \citenamefont {Van~der Avoird}}]{bukowski2007predictions}%
  \BibitemOpen
  \bibfield  {author} {\bibinfo {author} {\bibfnamefont {R.}~\bibnamefont
  {Bukowski}}, \bibinfo {author} {\bibfnamefont {K.}~\bibnamefont {Szalewicz}},
  \bibinfo {author} {\bibfnamefont {G.~C.}\ \bibnamefont {Groenenboom}},\ and\
  \bibinfo {author} {\bibfnamefont {A.}~\bibnamefont {Van~der Avoird}},\
  }\bibfield  {title} {\enquote {\bibinfo {title} {Predictions of the
  properties of water from first principles},}\ }\href@noop {} {\bibfield
  {journal} {\bibinfo  {journal} {Science}\ }\textbf {\bibinfo {volume}
  {315}},\ \bibinfo {pages} {1249--1252} (\bibinfo {year} {2007})}\BibitemShut
  {NoStop}%
\bibitem [{\citenamefont {Shinoda}\ and\ \citenamefont
  {Shiga}(2005)}]{shinoda2005quantum}%
  \BibitemOpen
  \bibfield  {author} {\bibinfo {author} {\bibfnamefont {W.}~\bibnamefont
  {Shinoda}}\ and\ \bibinfo {author} {\bibfnamefont {M.}~\bibnamefont
  {Shiga}},\ }\bibfield  {title} {\enquote {\bibinfo {title} {Quantum
  simulation of the heat capacity of water},}\ }\href@noop {} {\bibfield
  {journal} {\bibinfo  {journal} {Phys. Rev. E}\ }\textbf {\bibinfo {volume}
  {71}},\ \bibinfo {pages} {041204} (\bibinfo {year} {2005})}\BibitemShut
  {NoStop}%
\bibitem [{\citenamefont {Shiga}\ and\ \citenamefont
  {Shinoda}(2005)}]{shiga2005calculation}%
  \BibitemOpen
  \bibfield  {author} {\bibinfo {author} {\bibfnamefont {M.}~\bibnamefont
  {Shiga}}\ and\ \bibinfo {author} {\bibfnamefont {W.}~\bibnamefont
  {Shinoda}},\ }\bibfield  {title} {\enquote {\bibinfo {title} {Calculation of
  heat capacities of light and heavy water by path-integral molecular
  dynamics},}\ }\href@noop {} {\bibfield  {journal} {\bibinfo  {journal} {J.
  Chem. Phys.}\ }\textbf {\bibinfo {volume} {123}},\ \bibinfo {pages} {134502}
  (\bibinfo {year} {2005})}\BibitemShut {NoStop}%
\bibitem [{\citenamefont {Paesani}\ \emph {et~al.}(2006)\citenamefont
  {Paesani}, \citenamefont {Zhang}, \citenamefont {Case}, \citenamefont
  {{Cheatham III}},\ and\ \citenamefont {Voth}}]{paesani2006accurate}%
  \BibitemOpen
  \bibfield  {author} {\bibinfo {author} {\bibfnamefont {F.}~\bibnamefont
  {Paesani}}, \bibinfo {author} {\bibfnamefont {W.}~\bibnamefont {Zhang}},
  \bibinfo {author} {\bibfnamefont {D.~A.}\ \bibnamefont {Case}}, \bibinfo
  {author} {\bibfnamefont {T.~E.}\ \bibnamefont {{Cheatham III}}},\ and\
  \bibinfo {author} {\bibfnamefont {G.~A.}\ \bibnamefont {Voth}},\ }\bibfield
  {title} {\enquote {\bibinfo {title} {An accurate and simple quantum model for
  liquid water},}\ }\href@noop {} {\bibfield  {journal} {\bibinfo  {journal}
  {J. Chem. Phys.}\ }\textbf {\bibinfo {volume} {125}},\ \bibinfo {pages}
  {184507} (\bibinfo {year} {2006})}\BibitemShut {NoStop}%
\bibitem [{\citenamefont {Ceriotti}, \citenamefont {Manolopoulos},\ and\
  \citenamefont {Parrinello}(2011)}]{ceriotti2011accelerating}%
  \BibitemOpen
  \bibfield  {author} {\bibinfo {author} {\bibfnamefont {M.}~\bibnamefont
  {Ceriotti}}, \bibinfo {author} {\bibfnamefont {D.~E.}\ \bibnamefont
  {Manolopoulos}},\ and\ \bibinfo {author} {\bibfnamefont {M.}~\bibnamefont
  {Parrinello}},\ }\bibfield  {title} {\enquote {\bibinfo {title} {Accelerating
  the convergence of path integral dynamics with a generalized {L}angevin
  equation},}\ }\href@noop {} {\bibfield  {journal} {\bibinfo  {journal} {J.
  Chem. Phys.}\ }\textbf {\bibinfo {volume} {134}},\ \bibinfo {pages} {084104}
  (\bibinfo {year} {2011})}\BibitemShut {NoStop}%
\bibitem [{\citenamefont {Shvab}\ and\ \citenamefont
  {Sadus}(2016)}]{shvab2016atomistic}%
  \BibitemOpen
  \bibfield  {author} {\bibinfo {author} {\bibfnamefont {I.}~\bibnamefont
  {Shvab}}\ and\ \bibinfo {author} {\bibfnamefont {R.~J.}\ \bibnamefont
  {Sadus}},\ }\bibfield  {title} {\enquote {\bibinfo {title} {Atomistic water
  models: Aqueous thermodynamic properties from ambient to supercritical
  conditions},}\ }\href@noop {} {\bibfield  {journal} {\bibinfo  {journal}
  {Fluid Phase Equilibria}\ }\textbf {\bibinfo {volume} {407}},\ \bibinfo
  {pages} {7--30} (\bibinfo {year} {2016})}\BibitemShut {NoStop}%
\bibitem [{\citenamefont {Eltareb}, \citenamefont {Lopez},\ and\ \citenamefont
  {Giovambattista}(2021)}]{eltareb2021nuclear}%
  \BibitemOpen
  \bibfield  {author} {\bibinfo {author} {\bibfnamefont {A.}~\bibnamefont
  {Eltareb}}, \bibinfo {author} {\bibfnamefont {G.~E.}\ \bibnamefont {Lopez}},\
  and\ \bibinfo {author} {\bibfnamefont {N.}~\bibnamefont {Giovambattista}},\
  }\bibfield  {title} {\enquote {\bibinfo {title} {Nuclear quantum effects on
  the thermodynamic, structural, and dynamical properties of water},}\
  }\href@noop {} {\bibfield  {journal} {\bibinfo  {journal} {Phys. Chem. Chem.
  Phys.}\ }\textbf {\bibinfo {volume} {23}},\ \bibinfo {pages} {6914--6928}
  (\bibinfo {year} {2021})}\BibitemShut {NoStop}%
\bibitem [{\citenamefont {Parrinello}\ and\ \citenamefont
  {Rahman}(1984)}]{parrinello1984study}%
  \BibitemOpen
  \bibfield  {author} {\bibinfo {author} {\bibfnamefont {M.}~\bibnamefont
  {Parrinello}}\ and\ \bibinfo {author} {\bibfnamefont {A.}~\bibnamefont
  {Rahman}},\ }\bibfield  {title} {\enquote {\bibinfo {title} {Study of an f
  center in molten kcl},}\ }\href@noop {} {\bibfield  {journal} {\bibinfo
  {journal} {J. Chem. Phys.}\ }\textbf {\bibinfo {volume} {80}},\ \bibinfo
  {pages} {860--867} (\bibinfo {year} {1984})}\BibitemShut {NoStop}%
\bibitem [{\citenamefont {Hall}\ and\ \citenamefont
  {Berne}(1984)}]{hall1984nonergodicity}%
  \BibitemOpen
  \bibfield  {author} {\bibinfo {author} {\bibfnamefont {R.~W.}\ \bibnamefont
  {Hall}}\ and\ \bibinfo {author} {\bibfnamefont {B.~J.}\ \bibnamefont
  {Berne}},\ }\bibfield  {title} {\enquote {\bibinfo {title} {Nonergodicity in
  path integral molecular dynamics},}\ }\href@noop {} {\bibfield  {journal}
  {\bibinfo  {journal} {J. Chem. Phys.}\ }\textbf {\bibinfo {volume} {81}},\
  \bibinfo {pages} {3641--3643} (\bibinfo {year} {1984})}\BibitemShut {NoStop}%
\bibitem [{\citenamefont {Tuckerman}\ \emph {et~al.}(1993)\citenamefont
  {Tuckerman}, \citenamefont {Berne}, \citenamefont {Martyna},\ and\
  \citenamefont {Klein}}]{tuckerman1993efficient}%
  \BibitemOpen
  \bibfield  {author} {\bibinfo {author} {\bibfnamefont {M.~E.}\ \bibnamefont
  {Tuckerman}}, \bibinfo {author} {\bibfnamefont {B.~J.}\ \bibnamefont
  {Berne}}, \bibinfo {author} {\bibfnamefont {G.~J.}\ \bibnamefont {Martyna}},\
  and\ \bibinfo {author} {\bibfnamefont {M.~L.}\ \bibnamefont {Klein}},\
  }\bibfield  {title} {\enquote {\bibinfo {title} {Efficient molecular dynamics
  and hybrid {M}onte {C}arlo algorithms for path integrals},}\ }\href@noop {}
  {\bibfield  {journal} {\bibinfo  {journal} {J. Chem. Phys.}\ }\textbf
  {\bibinfo {volume} {99}},\ \bibinfo {pages} {2796--2808} (\bibinfo {year}
  {1993})}\BibitemShut {NoStop}%
\bibitem [{\citenamefont {Vega}\ \emph {et~al.}(2010)\citenamefont {Vega},
  \citenamefont {Conde}, \citenamefont {McBride}, \citenamefont {Abascal},
  \citenamefont {Noya}, \citenamefont {Ram{\'\i}rez},\ and\ \citenamefont
  {Ses{\'e}}}]{vega2010heat}%
  \BibitemOpen
  \bibfield  {author} {\bibinfo {author} {\bibfnamefont {C.}~\bibnamefont
  {Vega}}, \bibinfo {author} {\bibfnamefont {M.~M.}\ \bibnamefont {Conde}},
  \bibinfo {author} {\bibfnamefont {C.}~\bibnamefont {McBride}}, \bibinfo
  {author} {\bibfnamefont {J.~L.~F.}\ \bibnamefont {Abascal}}, \bibinfo
  {author} {\bibfnamefont {E.~G.}\ \bibnamefont {Noya}}, \bibinfo {author}
  {\bibfnamefont {R.}~\bibnamefont {Ram{\'\i}rez}},\ and\ \bibinfo {author}
  {\bibfnamefont {L.~M.}\ \bibnamefont {Ses{\'e}}},\ }\bibfield  {title}
  {\enquote {\bibinfo {title} {Heat capacity of water: A signature of nuclear
  quantum effects},}\ }\href@noop {} {\bibfield  {journal} {\bibinfo  {journal}
  {J. Chem. Phys.}\ }\textbf {\bibinfo {volume} {132}},\ \bibinfo {pages}
  {046101} (\bibinfo {year} {2010})}\BibitemShut {NoStop}%
\bibitem [{\citenamefont {Noya}\ \emph {et~al.}(2011)\citenamefont {Noya},
  \citenamefont {Ses{\'e}}, \citenamefont {Ram{\'\i}rez}, \citenamefont
  {McBride}, \citenamefont {Conde},\ and\ \citenamefont {Vega}}]{noya2011path}%
  \BibitemOpen
  \bibfield  {author} {\bibinfo {author} {\bibfnamefont {E.~G.}\ \bibnamefont
  {Noya}}, \bibinfo {author} {\bibfnamefont {L.~M.}\ \bibnamefont {Ses{\'e}}},
  \bibinfo {author} {\bibfnamefont {R.}~\bibnamefont {Ram{\'\i}rez}}, \bibinfo
  {author} {\bibfnamefont {C.}~\bibnamefont {McBride}}, \bibinfo {author}
  {\bibfnamefont {M.~M.}\ \bibnamefont {Conde}},\ and\ \bibinfo {author}
  {\bibfnamefont {C.}~\bibnamefont {Vega}},\ }\bibfield  {title} {\enquote
  {\bibinfo {title} {Path integral {M}onte {C}arlo simulations for rigid rotors
  and their application to water},}\ }\href@noop {} {\bibfield  {journal}
  {\bibinfo  {journal} {Mol. Phys.}\ }\textbf {\bibinfo {volume} {109}},\
  \bibinfo {pages} {149--168} (\bibinfo {year} {2011})}\BibitemShut {NoStop}%
\bibitem [{\citenamefont {Berta}\ \emph {et~al.}(2020)\citenamefont {Berta},
  \citenamefont {Ferenc}, \citenamefont {Bak{\'o}},\ and\ \citenamefont
  {Madar{\'a}sz}}]{berta2020nuclear}%
  \BibitemOpen
  \bibfield  {author} {\bibinfo {author} {\bibfnamefont {D.}~\bibnamefont
  {Berta}}, \bibinfo {author} {\bibfnamefont {D.}~\bibnamefont {Ferenc}},
  \bibinfo {author} {\bibfnamefont {I.}~\bibnamefont {Bak{\'o}}},\ and\
  \bibinfo {author} {\bibfnamefont {{\'A}.}~\bibnamefont {Madar{\'a}sz}},\
  }\bibfield  {title} {\enquote {\bibinfo {title} {Nuclear quantum effects from
  the analysis of smoothed trajectories: Pilot study for water},}\ }\href@noop
  {} {\bibfield  {journal} {\bibinfo  {journal} {J. Chem. Theory Comput.}\
  }\textbf {\bibinfo {volume} {16}},\ \bibinfo {pages} {3316--3334} (\bibinfo
  {year} {2020})}\BibitemShut {NoStop}%
\bibitem [{\citenamefont {Savoia}\ \emph {et~al.}(2025)\citenamefont {Savoia},
  \citenamefont {Oyarzua}, \citenamefont {Todd},\ and\ \citenamefont
  {Sadus}}]{savoia2025influence}%
  \BibitemOpen
  \bibfield  {author} {\bibinfo {author} {\bibfnamefont {E.}~\bibnamefont
  {Savoia}}, \bibinfo {author} {\bibfnamefont {E.}~\bibnamefont {Oyarzua}},
  \bibinfo {author} {\bibfnamefont {B.~D.}\ \bibnamefont {Todd}},\ and\
  \bibinfo {author} {\bibfnamefont {R.~J.}\ \bibnamefont {Sadus}},\ }\bibfield
  {title} {\enquote {\bibinfo {title} {Influence of quantum corrections on the
  predicted isobaric heat capacity of polarizable water models},}\ }\href@noop
  {} {\bibfield  {journal} {\bibinfo  {journal} {J. Chem. Phys.}\ }\textbf
  {\bibinfo {volume} {162}},\ \bibinfo {pages} {144503} (\bibinfo {year}
  {2025})}\BibitemShut {NoStop}%
\bibitem [{\citenamefont {Behler}\ and\ \citenamefont
  {Parrinello}(2007)}]{behler2007generalized}%
  \BibitemOpen
  \bibfield  {author} {\bibinfo {author} {\bibfnamefont {J.}~\bibnamefont
  {Behler}}\ and\ \bibinfo {author} {\bibfnamefont {M.}~\bibnamefont
  {Parrinello}},\ }\bibfield  {title} {\enquote {\bibinfo {title} {Generalized
  neural-network representation of high-dimensional potential-energy
  surfaces},}\ }\href@noop {} {\bibfield  {journal} {\bibinfo  {journal} {Phys.
  Rev. Lett.}\ }\textbf {\bibinfo {volume} {98}},\ \bibinfo {pages} {146401}
  (\bibinfo {year} {2007})}\BibitemShut {NoStop}%
\bibitem [{\citenamefont {Behler}(2011)}]{behler2011neural}%
  \BibitemOpen
  \bibfield  {author} {\bibinfo {author} {\bibfnamefont {J.}~\bibnamefont
  {Behler}},\ }\bibfield  {title} {\enquote {\bibinfo {title} {Neural network
  potential-energy surfaces in chemistry: a tool for large-scale
  simulations},}\ }\href@noop {} {\bibfield  {journal} {\bibinfo  {journal}
  {Phys. Chem. Chem. Phys.}\ }\textbf {\bibinfo {volume} {13}},\ \bibinfo
  {pages} {17930--17955} (\bibinfo {year} {2011})}\BibitemShut {NoStop}%
\bibitem [{\citenamefont {Behler}(2015)}]{behler2015constructing}%
  \BibitemOpen
  \bibfield  {author} {\bibinfo {author} {\bibfnamefont {J.}~\bibnamefont
  {Behler}},\ }\bibfield  {title} {\enquote {\bibinfo {title} {Constructing
  high-dimensional neural network potentials: A tutorial review},}\ }\href@noop
  {} {\bibfield  {journal} {\bibinfo  {journal} {Int. J. Quant. Chem.}\
  }\textbf {\bibinfo {volume} {115}},\ \bibinfo {pages} {1032--1050} (\bibinfo
  {year} {2015})}\BibitemShut {NoStop}%
\bibitem [{\citenamefont {Behler}(2016)}]{behler2016perspective}%
  \BibitemOpen
  \bibfield  {author} {\bibinfo {author} {\bibfnamefont {J.}~\bibnamefont
  {Behler}},\ }\bibfield  {title} {\enquote {\bibinfo {title} {Perspective:
  Machine learning potentials for atomistic simulations},}\ }\href@noop {}
  {\bibfield  {journal} {\bibinfo  {journal} {J. Chem. Phys.}\ }\textbf
  {\bibinfo {volume} {145}},\ \bibinfo {pages} {170901} (\bibinfo {year}
  {2016})}\BibitemShut {NoStop}%
\bibitem [{\citenamefont {Behler}(2021)}]{behler2021four}%
  \BibitemOpen
  \bibfield  {author} {\bibinfo {author} {\bibfnamefont {J.}~\bibnamefont
  {Behler}},\ }\bibfield  {title} {\enquote {\bibinfo {title} {Four generations
  of high-dimensional neural network potentials},}\ }\href@noop {} {\bibfield
  {journal} {\bibinfo  {journal} {Chem. Rev.}\ }\textbf {\bibinfo {volume}
  {121}},\ \bibinfo {pages} {10037--10072} (\bibinfo {year}
  {2021})}\BibitemShut {NoStop}%
\bibitem [{\citenamefont {Marx}\ and\ \citenamefont
  {Parrinello}(1996)}]{marx1996ab}%
  \BibitemOpen
  \bibfield  {author} {\bibinfo {author} {\bibfnamefont {D.}~\bibnamefont
  {Marx}}\ and\ \bibinfo {author} {\bibfnamefont {M.}~\bibnamefont
  {Parrinello}},\ }\bibfield  {title} {\enquote {\bibinfo {title} {Ab initio
  path integral molecular dynamics: Basic ideas},}\ }\href@noop {} {\bibfield
  {journal} {\bibinfo  {journal} {J. Chem. Phys.}\ }\textbf {\bibinfo {volume}
  {104}},\ \bibinfo {pages} {4077--4082} (\bibinfo {year} {1996})}\BibitemShut
  {NoStop}%
\bibitem [{\citenamefont {Shiga}, \citenamefont {Tachikawa},\ and\
  \citenamefont {Miura}(2001)}]{shiga2001unified}%
  \BibitemOpen
  \bibfield  {author} {\bibinfo {author} {\bibfnamefont {M.}~\bibnamefont
  {Shiga}}, \bibinfo {author} {\bibfnamefont {M.}~\bibnamefont {Tachikawa}},\
  and\ \bibinfo {author} {\bibfnamefont {S.}~\bibnamefont {Miura}},\ }\bibfield
   {title} {\enquote {\bibinfo {title} {A unified scheme for ab initio
  molecular orbital theory and path integral molecular dynamics},}\ }\href@noop
  {} {\bibfield  {journal} {\bibinfo  {journal} {J. Chem. Phys.}\ }\textbf
  {\bibinfo {volume} {115}},\ \bibinfo {pages} {9149--9159} (\bibinfo {year}
  {2001})}\BibitemShut {NoStop}%
\bibitem [{\citenamefont {Cheng}, \citenamefont {Behler},\ and\ \citenamefont
  {Ceriotti}(2016)}]{ChengHDNNP2016}%
  \BibitemOpen
  \bibfield  {author} {\bibinfo {author} {\bibfnamefont {B.}~\bibnamefont
  {Cheng}}, \bibinfo {author} {\bibfnamefont {J.}~\bibnamefont {Behler}},\ and\
  \bibinfo {author} {\bibfnamefont {M.}~\bibnamefont {Ceriotti}},\ }\bibfield
  {title} {\enquote {\bibinfo {title} {Nuclear quantum effects in water at the
  triple point: Using theory as a link between experiments},}\ }\href
  {https://doi.org/10.1021/acs.jpclett.6b00729} {\bibfield  {journal} {\bibinfo
   {journal} {J. Phys. Chem. Lett.}\ }\textbf {\bibinfo {volume} {7}},\
  \bibinfo {pages} {2210--2215} (\bibinfo {year} {2016})}\BibitemShut {NoStop}%
\bibitem [{\citenamefont {Kapil}, \citenamefont {Behler},\ and\ \citenamefont
  {Ceriotti}(2016)}]{P4971}%
  \BibitemOpen
  \bibfield  {author} {\bibinfo {author} {\bibfnamefont {V.}~\bibnamefont
  {Kapil}}, \bibinfo {author} {\bibfnamefont {J.}~\bibnamefont {Behler}},\ and\
  \bibinfo {author} {\bibfnamefont {M.}~\bibnamefont {Ceriotti}},\ }\bibfield
  {title} {\enquote {\bibinfo {title} {High order path integrals made easy},}\
  }\href@noop {} {\bibfield  {journal} {\bibinfo  {journal} {J. Chem. Phys.}\
  }\textbf {\bibinfo {volume} {145}},\ \bibinfo {pages} {234103} (\bibinfo
  {year} {2016})}\BibitemShut {NoStop}%
\bibitem [{\citenamefont {Hellstr{\"o}m}, \citenamefont {Ceriotti},\ and\
  \citenamefont {Behler}(2018)}]{P5631}%
  \BibitemOpen
  \bibfield  {author} {\bibinfo {author} {\bibfnamefont {M.}~\bibnamefont
  {Hellstr{\"o}m}}, \bibinfo {author} {\bibfnamefont {M.}~\bibnamefont
  {Ceriotti}},\ and\ \bibinfo {author} {\bibfnamefont {J.}~\bibnamefont
  {Behler}},\ }\bibfield  {title} {\enquote {\bibinfo {title} {Nuclear quantum
  effects in sodium hydroxide solutions from neural network molecular dynamics
  simulations},}\ }\href@noop {} {\bibfield  {journal} {\bibinfo  {journal} {J.
  Phys. Chem. B}\ }\textbf {\bibinfo {volume} {122}},\ \bibinfo {pages}
  {10158--10171} (\bibinfo {year} {2018})}\BibitemShut {NoStop}%
\bibitem [{\citenamefont {Cheng}\ \emph {et~al.}(2019)\citenamefont {Cheng},
  \citenamefont {Engel}, \citenamefont {Behler}, \citenamefont {Dellago},\ and\
  \citenamefont {Ceriotti}}]{ChengHDNNP2019}%
  \BibitemOpen
  \bibfield  {author} {\bibinfo {author} {\bibfnamefont {B.}~\bibnamefont
  {Cheng}}, \bibinfo {author} {\bibfnamefont {E.~A.}\ \bibnamefont {Engel}},
  \bibinfo {author} {\bibfnamefont {J.}~\bibnamefont {Behler}}, \bibinfo
  {author} {\bibfnamefont {C.}~\bibnamefont {Dellago}},\ and\ \bibinfo {author}
  {\bibfnamefont {M.}~\bibnamefont {Ceriotti}},\ }\bibfield  {title} {\enquote
  {\bibinfo {title} {Ab initio thermodynamics of liquid and solid water},}\
  }\href {https://doi.org/10.1073/pnas.1815117116} {\bibfield  {journal}
  {\bibinfo  {journal} {Proc. Natl. Acad. Sci. U.S.A.}\ }\textbf {\bibinfo
  {volume} {116}},\ \bibinfo {pages} {1110--1115} (\bibinfo {year}
  {2019})}\BibitemShut {NoStop}%
\bibitem [{\citenamefont {Reinhardt}\ and\ \citenamefont
  {Cheng}(2021)}]{ReinhardtHDNNP2021}%
  \BibitemOpen
  \bibfield  {author} {\bibinfo {author} {\bibfnamefont {A.}~\bibnamefont
  {Reinhardt}}\ and\ \bibinfo {author} {\bibfnamefont {B.}~\bibnamefont
  {Cheng}},\ }\bibfield  {title} {\enquote {\bibinfo {title}
  {Quantum-mechanical exploration of the phase diagram of water},}\ }\href
  {https://doi.org/10.1038/s41467-020-20821-w} {\bibfield  {journal} {\bibinfo
  {journal} {Nat. Commun.}\ }\textbf {\bibinfo {volume} {12}},\ \bibinfo
  {pages} {588} (\bibinfo {year} {2021})}\BibitemShut {NoStop}%
\bibitem [{\citenamefont {Daru}\ \emph {et~al.}(2022)\citenamefont {Daru},
  \citenamefont {Forbert}, \citenamefont {Behler},\ and\ \citenamefont
  {Marx}}]{daru2022coupled}%
  \BibitemOpen
  \bibfield  {author} {\bibinfo {author} {\bibfnamefont {J.}~\bibnamefont
  {Daru}}, \bibinfo {author} {\bibfnamefont {H.}~\bibnamefont {Forbert}},
  \bibinfo {author} {\bibfnamefont {J.}~\bibnamefont {Behler}},\ and\ \bibinfo
  {author} {\bibfnamefont {D.}~\bibnamefont {Marx}},\ }\bibfield  {title}
  {\enquote {\bibinfo {title} {Coupled cluster molecular dynamics of condensed
  phase systems enabled by machine learning potentials: Liquid water
  benchmark},}\ }\href@noop {} {\bibfield  {journal} {\bibinfo  {journal}
  {Phys. Rev. Lett.}\ }\textbf {\bibinfo {volume} {129}},\ \bibinfo {pages}
  {226001} (\bibinfo {year} {2022})}\BibitemShut {NoStop}%
\bibitem [{\citenamefont {Thomsen}\ \emph {et~al.}(2024)\citenamefont
  {Thomsen}, \citenamefont {Nagai}, \citenamefont {Kobayashi}, \citenamefont
  {Hamada},\ and\ \citenamefont {Shiga}}]{thomsen2024self}%
  \BibitemOpen
  \bibfield  {author} {\bibinfo {author} {\bibfnamefont {B.}~\bibnamefont
  {Thomsen}}, \bibinfo {author} {\bibfnamefont {Y.}~\bibnamefont {Nagai}},
  \bibinfo {author} {\bibfnamefont {K.}~\bibnamefont {Kobayashi}}, \bibinfo
  {author} {\bibfnamefont {I.}~\bibnamefont {Hamada}},\ and\ \bibinfo {author}
  {\bibfnamefont {M.}~\bibnamefont {Shiga}},\ }\bibfield  {title} {\enquote
  {\bibinfo {title} {Self-learning path integral hybrid {M}onte {C}arlo with
  mixed ab initio and machine learning potentials for modeling nuclear quantum
  effects in water},}\ }\href@noop {} {\bibfield  {journal} {\bibinfo
  {journal} {J. Chem. Phys.}\ }\textbf {\bibinfo {volume} {161}},\ \bibinfo
  {pages} {204109} (\bibinfo {year} {2024})}\BibitemShut {NoStop}%
\bibitem [{\citenamefont {Stolte}\ \emph {et~al.}(2024)\citenamefont {Stolte},
  \citenamefont {Daru}, \citenamefont {Forbert}, \citenamefont {Behler},\ and\
  \citenamefont {Marx}}]{stolte2024nuclear}%
  \BibitemOpen
  \bibfield  {author} {\bibinfo {author} {\bibfnamefont {N.}~\bibnamefont
  {Stolte}}, \bibinfo {author} {\bibfnamefont {J.}~\bibnamefont {Daru}},
  \bibinfo {author} {\bibfnamefont {H.}~\bibnamefont {Forbert}}, \bibinfo
  {author} {\bibfnamefont {J.}~\bibnamefont {Behler}},\ and\ \bibinfo {author}
  {\bibfnamefont {D.}~\bibnamefont {Marx}},\ }\bibfield  {title} {\enquote
  {\bibinfo {title} {Nuclear quantum effects in liquid water are marginal for
  its average structure but significant for dynamics},}\ }\href@noop {}
  {\bibfield  {journal} {\bibinfo  {journal} {J. Phys. Chem. Lett.}\ }\textbf
  {\bibinfo {volume} {15}},\ \bibinfo {pages} {12144--12150} (\bibinfo {year}
  {2024})}\BibitemShut {NoStop}%
\bibitem [{\citenamefont {Stolte}\ \emph {et~al.}(2025)\citenamefont {Stolte},
  \citenamefont {Daru}, \citenamefont {Forbert}, \citenamefont {Marx},\ and\
  \citenamefont {Behler}}]{stolte2025random}%
  \BibitemOpen
  \bibfield  {author} {\bibinfo {author} {\bibfnamefont {N.}~\bibnamefont
  {Stolte}}, \bibinfo {author} {\bibfnamefont {J.}~\bibnamefont {Daru}},
  \bibinfo {author} {\bibfnamefont {H.}~\bibnamefont {Forbert}}, \bibinfo
  {author} {\bibfnamefont {D.}~\bibnamefont {Marx}},\ and\ \bibinfo {author}
  {\bibfnamefont {J.}~\bibnamefont {Behler}},\ }\bibfield  {title} {\enquote
  {\bibinfo {title} {Random sampling versus active learning algorithms for
  machine learning potentials of quantum liquid water},}\ }\href@noop {}
  {\bibfield  {journal} {\bibinfo  {journal} {J. Chem. Theory Comput.}\
  }\textbf {\bibinfo {volume} {21}},\ \bibinfo {pages} {886--899} (\bibinfo
  {year} {2025})}\BibitemShut {NoStop}%
\bibitem [{\citenamefont {Malik}\ \emph {et~al.}(2025)\citenamefont {Malik},
  \citenamefont {Stolte}, \citenamefont {Forbert}, \citenamefont {Chandra},\
  and\ \citenamefont {Marx}}]{malik2025accurate}%
  \BibitemOpen
  \bibfield  {author} {\bibinfo {author} {\bibfnamefont {R.}~\bibnamefont
  {Malik}}, \bibinfo {author} {\bibfnamefont {N.}~\bibnamefont {Stolte}},
  \bibinfo {author} {\bibfnamefont {H.}~\bibnamefont {Forbert}}, \bibinfo
  {author} {\bibfnamefont {A.}~\bibnamefont {Chandra}},\ and\ \bibinfo {author}
  {\bibfnamefont {D.}~\bibnamefont {Marx}},\ }\bibfield  {title} {\enquote
  {\bibinfo {title} {Accurate determination of isotope effects on the dynamics
  of h-bond breaking and making in liquid water},}\ }\href@noop {} {\bibfield
  {journal} {\bibinfo  {journal} {J. Phys. Chem. Lett.}\ }\textbf {\bibinfo
  {volume} {16}},\ \bibinfo {pages} {3727--3733} (\bibinfo {year}
  {2025})}\BibitemShut {NoStop}%
\bibitem [{\citenamefont {Omranpour}\ \emph {et~al.}(2024)\citenamefont
  {Omranpour}, \citenamefont {Montero De~Hijes}, \citenamefont {Behler},\ and\
  \citenamefont {Dellago}}]{OmranpourReview}%
  \BibitemOpen
  \bibfield  {author} {\bibinfo {author} {\bibfnamefont {A.}~\bibnamefont
  {Omranpour}}, \bibinfo {author} {\bibfnamefont {P.}~\bibnamefont {Montero
  De~Hijes}}, \bibinfo {author} {\bibfnamefont {J.}~\bibnamefont {Behler}},\
  and\ \bibinfo {author} {\bibfnamefont {C.}~\bibnamefont {Dellago}},\
  }\bibfield  {title} {\enquote {\bibinfo {title} {Perspective: Atomistic
  simulations of water and aqueous systems with machine learning potentials},}\
  }\href {https://doi.org/10.1063/5.0201241} {\bibfield  {journal} {\bibinfo
  {journal} {J. Chem. Phys.}\ }\textbf {\bibinfo {volume} {160}},\ \bibinfo
  {pages} {170901} (\bibinfo {year} {2024})}\BibitemShut {NoStop}%
\bibitem [{\citenamefont {Ruiz-Barragan}, \citenamefont {Ishimura},\ and\
  \citenamefont {Shiga}(2016)}]{ruiz2016hierarchical}%
  \BibitemOpen
  \bibfield  {author} {\bibinfo {author} {\bibfnamefont {S.}~\bibnamefont
  {Ruiz-Barragan}}, \bibinfo {author} {\bibfnamefont {K.}~\bibnamefont
  {Ishimura}},\ and\ \bibinfo {author} {\bibfnamefont {M.}~\bibnamefont
  {Shiga}},\ }\bibfield  {title} {\enquote {\bibinfo {title} {On the
  hierarchical parallelization of ab initio simulations},}\ }\href@noop {}
  {\bibfield  {journal} {\bibinfo  {journal} {Chem. Phys. Lett.}\ }\textbf
  {\bibinfo {volume} {646}},\ \bibinfo {pages} {130--135} (\bibinfo {year}
  {2016})}\BibitemShut {NoStop}%
\bibitem [{\citenamefont {Cao}\ and\ \citenamefont
  {Martyna}(1996)}]{cao1996adiabatic}%
  \BibitemOpen
  \bibfield  {author} {\bibinfo {author} {\bibfnamefont {J.}~\bibnamefont
  {Cao}}\ and\ \bibinfo {author} {\bibfnamefont {G.~J.}\ \bibnamefont
  {Martyna}},\ }\bibfield  {title} {\enquote {\bibinfo {title} {Adiabatic path
  integral molecular dynamics methods. {II}. algorithms},}\ }\href@noop {}
  {\bibfield  {journal} {\bibinfo  {journal} {J. Chem. Phys.}\ }\textbf
  {\bibinfo {volume} {104}},\ \bibinfo {pages} {2028--2035} (\bibinfo {year}
  {1996})}\BibitemShut {NoStop}%
\bibitem [{\citenamefont {Nos{\'e}}(1984)}]{nose1984unified}%
  \BibitemOpen
  \bibfield  {author} {\bibinfo {author} {\bibfnamefont {S.}~\bibnamefont
  {Nos{\'e}}},\ }\bibfield  {title} {\enquote {\bibinfo {title} {A unified
  formulation of the constant temperature molecular dynamics methods},}\
  }\href@noop {} {\bibfield  {journal} {\bibinfo  {journal} {J. Chem. Phys.}\
  }\textbf {\bibinfo {volume} {81}},\ \bibinfo {pages} {511--519} (\bibinfo
  {year} {1984})}\BibitemShut {NoStop}%
\bibitem [{\citenamefont {Hoover}(1985)}]{hoover1985canonical}%
  \BibitemOpen
  \bibfield  {author} {\bibinfo {author} {\bibfnamefont {W.~G.}\ \bibnamefont
  {Hoover}},\ }\bibfield  {title} {\enquote {\bibinfo {title} {Canonical
  dynamics: Equilibrium phase-space distributions},}\ }\href@noop {} {\bibfield
   {journal} {\bibinfo  {journal} {Phys. Rev. A}\ }\textbf {\bibinfo {volume}
  {31}},\ \bibinfo {pages} {1695} (\bibinfo {year} {1985})}\BibitemShut
  {NoStop}%
\bibitem [{\citenamefont {Martyna}, \citenamefont {Klein},\ and\ \citenamefont
  {Tuckerman}(1992)}]{martyna1992nose}%
  \BibitemOpen
  \bibfield  {author} {\bibinfo {author} {\bibfnamefont {G.~J.}\ \bibnamefont
  {Martyna}}, \bibinfo {author} {\bibfnamefont {M.~L.}\ \bibnamefont {Klein}},\
  and\ \bibinfo {author} {\bibfnamefont {M.}~\bibnamefont {Tuckerman}},\
  }\bibfield  {title} {\enquote {\bibinfo {title} {Nos{\'e}--{H}oover chains:
  The canonical ensemble via continuous dynamics},}\ }\href@noop {} {\bibfield
  {journal} {\bibinfo  {journal} {J. Chem. Phys.}\ }\textbf {\bibinfo {volume}
  {97}},\ \bibinfo {pages} {2635--2643} (\bibinfo {year} {1992})}\BibitemShut
  {NoStop}%
\bibitem [{\citenamefont {Marx}\ and\ \citenamefont
  {Hutter}(2009)}]{marx2009ab}%
  \BibitemOpen
  \bibfield  {author} {\bibinfo {author} {\bibfnamefont {D.}~\bibnamefont
  {Marx}}\ and\ \bibinfo {author} {\bibfnamefont {J.}~\bibnamefont {Hutter}},\
  }\href@noop {} {\emph {\bibinfo {title} {Ab initio molecular dynamics: basic
  theory and advanced methods}}}\ (\bibinfo  {publisher} {Cambridge University
  Press},\ \bibinfo {year} {2009})\BibitemShut {NoStop}%
\bibitem [{\citenamefont {Tuckerman}(2010)}]{tuckerman2010statistical}%
  \BibitemOpen
  \bibfield  {author} {\bibinfo {author} {\bibfnamefont {M.~E.}\ \bibnamefont
  {Tuckerman}},\ }\href@noop {} {\emph {\bibinfo {title} {Statistical
  mechanics: theory and molecular simulation}}}\ (\bibinfo  {publisher} {Oxford
  University Press},\ \bibinfo {year} {2010})\BibitemShut {NoStop}%
\bibitem [{\citenamefont {Shiga}(2018)}]{shiga2018path}%
  \BibitemOpen
  \bibfield  {author} {\bibinfo {author} {\bibfnamefont {M.}~\bibnamefont
  {Shiga}},\ }\bibfield  {title} {\enquote {\bibinfo {title} {Path integral
  simulations},}\ }in\ \href@noop {} {\emph {\bibinfo {booktitle} {Reference
  Module in Chemistry, Molecular Sciences and Chemical Engineering}}}\
  (\bibinfo  {publisher} {Elsevier},\ \bibinfo {year} {2018})\BibitemShut
  {NoStop}%
\bibitem [{\citenamefont {Tuckerman}, \citenamefont {Berne},\ and\
  \citenamefont {Martyna}(1992)}]{tuckerman1992reversible}%
  \BibitemOpen
  \bibfield  {author} {\bibinfo {author} {\bibfnamefont {M.}~\bibnamefont
  {Tuckerman}}, \bibinfo {author} {\bibfnamefont {B.~J.}\ \bibnamefont
  {Berne}},\ and\ \bibinfo {author} {\bibfnamefont {G.~J.}\ \bibnamefont
  {Martyna}},\ }\bibfield  {title} {\enquote {\bibinfo {title} {Reversible
  multiple time scale molecular dynamics},}\ }\href@noop {} {\bibfield
  {journal} {\bibinfo  {journal} {J. Chem. Phys.}\ }\textbf {\bibinfo {volume}
  {97}},\ \bibinfo {pages} {1990--2001} (\bibinfo {year} {1992})}\BibitemShut
  {NoStop}%
\bibitem [{\citenamefont {Cao}\ and\ \citenamefont
  {Voth}(1994)}]{cao1994formulation}%
  \BibitemOpen
  \bibfield  {author} {\bibinfo {author} {\bibfnamefont {J.}~\bibnamefont
  {Cao}}\ and\ \bibinfo {author} {\bibfnamefont {G.~A.}\ \bibnamefont {Voth}},\
  }\bibfield  {title} {\enquote {\bibinfo {title} {The formulation of quantum
  statistical mechanics based on the {F}eynman path centroid density. {II}.
  dynamical properties},}\ }\href@noop {} {\bibfield  {journal} {\bibinfo
  {journal} {J. Chem. Phys.}\ }\textbf {\bibinfo {volume} {100}},\ \bibinfo
  {pages} {5106--5117} (\bibinfo {year} {1994})}\BibitemShut {NoStop}%
\bibitem [{\citenamefont {Craig}\ and\ \citenamefont
  {Manolopoulos}(2004)}]{craig2004quantum}%
  \BibitemOpen
  \bibfield  {author} {\bibinfo {author} {\bibfnamefont {I.~R.}\ \bibnamefont
  {Craig}}\ and\ \bibinfo {author} {\bibfnamefont {D.~E.}\ \bibnamefont
  {Manolopoulos}},\ }\bibfield  {title} {\enquote {\bibinfo {title} {Quantum
  statistics and classical mechanics: Real time correlation functions from ring
  polymer molecular dynamics},}\ }\href@noop {} {\bibfield  {journal} {\bibinfo
   {journal} {J. Chem. Phys.}\ }\textbf {\bibinfo {volume} {121}},\ \bibinfo
  {pages} {3368--3373} (\bibinfo {year} {2004})}\BibitemShut {NoStop}%
\bibitem [{\citenamefont {Nagai}\ \emph {et~al.}(2020)\citenamefont {Nagai},
  \citenamefont {Okumura}, \citenamefont {Kobayashi},\ and\ \citenamefont
  {Shiga}}]{nagai2020self}%
  \BibitemOpen
  \bibfield  {author} {\bibinfo {author} {\bibfnamefont {Y.}~\bibnamefont
  {Nagai}}, \bibinfo {author} {\bibfnamefont {M.}~\bibnamefont {Okumura}},
  \bibinfo {author} {\bibfnamefont {K.}~\bibnamefont {Kobayashi}},\ and\
  \bibinfo {author} {\bibfnamefont {M.}~\bibnamefont {Shiga}},\ }\bibfield
  {title} {\enquote {\bibinfo {title} {Self-learning hybrid {M}onte {C}arlo: A
  first-principles approach},}\ }\href@noop {} {\bibfield  {journal} {\bibinfo
  {journal} {Phys. Rev. B}\ }\textbf {\bibinfo {volume} {102}},\ \bibinfo
  {pages} {041124} (\bibinfo {year} {2020})}\BibitemShut {NoStop}%
\bibitem [{\citenamefont {Glaesemann}\ and\ \citenamefont
  {Fried}(2002)}]{glaesemann2002improved}%
  \BibitemOpen
  \bibfield  {author} {\bibinfo {author} {\bibfnamefont {K.~R.}\ \bibnamefont
  {Glaesemann}}\ and\ \bibinfo {author} {\bibfnamefont {L.~E.}\ \bibnamefont
  {Fried}},\ }\bibfield  {title} {\enquote {\bibinfo {title} {Improved heat
  capacity estimator for path integral simulations},}\ }\href@noop {}
  {\bibfield  {journal} {\bibinfo  {journal} {J. Chem. Phys.}\ }\textbf
  {\bibinfo {volume} {117}},\ \bibinfo {pages} {3020--3026} (\bibinfo {year}
  {2002})}\BibitemShut {NoStop}%
\bibitem [{\citenamefont {Predescu}\ \emph {et~al.}(2003)\citenamefont
  {Predescu}, \citenamefont {Sabo}, \citenamefont {Doll},\ and\ \citenamefont
  {Freeman}}]{predescu2003heat}%
  \BibitemOpen
  \bibfield  {author} {\bibinfo {author} {\bibfnamefont {C.}~\bibnamefont
  {Predescu}}, \bibinfo {author} {\bibfnamefont {D.}~\bibnamefont {Sabo}},
  \bibinfo {author} {\bibfnamefont {J.}~\bibnamefont {Doll}},\ and\ \bibinfo
  {author} {\bibfnamefont {D.~L.}\ \bibnamefont {Freeman}},\ }\bibfield
  {title} {\enquote {\bibinfo {title} {Heat capacity estimators for random
  series path-integral methods by finite-difference schemes},}\ }\href@noop {}
  {\bibfield  {journal} {\bibinfo  {journal} {J. Chem. Phys.}\ }\textbf
  {\bibinfo {volume} {119}},\ \bibinfo {pages} {12119--12128} (\bibinfo {year}
  {2003})}\BibitemShut {NoStop}%
\bibitem [{\citenamefont {Yamamoto}(2005)}]{yamamoto2005path}%
  \BibitemOpen
  \bibfield  {author} {\bibinfo {author} {\bibfnamefont {T.~M.}\ \bibnamefont
  {Yamamoto}},\ }\bibfield  {title} {\enquote {\bibinfo {title} {Path-integral
  virial estimator based on the scaling of fluctuation coordinates: Application
  to quantum clusters with fourth-order propagators},}\ }\href@noop {}
  {\bibfield  {journal} {\bibinfo  {journal} {J. Chem. Phys.}\ }\textbf
  {\bibinfo {volume} {123}},\ \bibinfo {pages} {104101} (\bibinfo {year}
  {2005})}\BibitemShut {NoStop}%
\bibitem [{\citenamefont {Marienhagen}\ and\ \citenamefont
  {Meier}(2024)}]{marienhagen2024calculation}%
  \BibitemOpen
  \bibfield  {author} {\bibinfo {author} {\bibfnamefont {P.}~\bibnamefont
  {Marienhagen}}\ and\ \bibinfo {author} {\bibfnamefont {K.}~\bibnamefont
  {Meier}},\ }\bibfield  {title} {\enquote {\bibinfo {title} {Calculation of
  thermodynamic properties of helium using path integral {M}onte {C}arlo
  simulations in the {N}p{T} ensemble and ab initio potentials},}\ }\href@noop
  {} {\bibfield  {journal} {\bibinfo  {journal} {J. Chem. Phys.}\ }\textbf
  {\bibinfo {volume} {161}} (\bibinfo {year} {2024})}\BibitemShut {NoStop}%
\bibitem [{\citenamefont {Moustafa}\ and\ \citenamefont
  {Schultz}(2024)}]{moustafa2024generalized}%
  \BibitemOpen
  \bibfield  {author} {\bibinfo {author} {\bibfnamefont {S.~G.}\ \bibnamefont
  {Moustafa}}\ and\ \bibinfo {author} {\bibfnamefont {A.~J.}\ \bibnamefont
  {Schultz}},\ }\bibfield  {title} {\enquote {\bibinfo {title} {Generalized
  path integral energy and heat capacity estimators of quantum oscillators and
  crystals using harmonic mapping},}\ }\href@noop {} {\bibfield  {journal}
  {\bibinfo  {journal} {J. Chem. Theory Comput.}\ }\textbf {\bibinfo {volume}
  {20}},\ \bibinfo {pages} {10132--10146} (\bibinfo {year} {2024})}\BibitemShut
  {NoStop}%
\bibitem [{\citenamefont {Sadus}(2023)}]{sadus2023molecular}%
  \BibitemOpen
  \bibfield  {author} {\bibinfo {author} {\bibfnamefont {R.~J.}\ \bibnamefont
  {Sadus}},\ }\bibfield  {title} {\enquote {\bibinfo {title} {Molecular
  simulation meets machine learning},}\ }\href@noop {} {\bibfield  {journal}
  {\bibinfo  {journal} {J. Chem. Eng. Data}\ }\textbf {\bibinfo {volume}
  {69}},\ \bibinfo {pages} {3--11} (\bibinfo {year} {2023})}\BibitemShut
  {NoStop}%
\bibitem [{\citenamefont {Kobayashi}\ \emph {et~al.}(2021)\citenamefont
  {Kobayashi}, \citenamefont {Nagai}, \citenamefont {Itakura},\ and\
  \citenamefont {Shiga}}]{kobayashi2021self}%
  \BibitemOpen
  \bibfield  {author} {\bibinfo {author} {\bibfnamefont {K.}~\bibnamefont
  {Kobayashi}}, \bibinfo {author} {\bibfnamefont {Y.}~\bibnamefont {Nagai}},
  \bibinfo {author} {\bibfnamefont {M.}~\bibnamefont {Itakura}},\ and\ \bibinfo
  {author} {\bibfnamefont {M.}~\bibnamefont {Shiga}},\ }\bibfield  {title}
  {\enquote {\bibinfo {title} {Self-learning hybrid monte carlo method for
  isothermal--isobaric ensemble: Application to liquid silica},}\ }\href@noop
  {} {\bibfield  {journal} {\bibinfo  {journal} {The Journal of Chemical
  Physics}\ }\textbf {\bibinfo {volume} {155}},\ \bibinfo {pages} {034106}
  (\bibinfo {year} {2021})}\BibitemShut {NoStop}%
\bibitem [{\citenamefont {Quaranta}, \citenamefont {Hellstr{\"o}m},\ and\
  \citenamefont {Behler}(2017)}]{quaranta2017}%
  \BibitemOpen
  \bibfield  {author} {\bibinfo {author} {\bibfnamefont {V.}~\bibnamefont
  {Quaranta}}, \bibinfo {author} {\bibfnamefont {M.}~\bibnamefont
  {Hellstr{\"o}m}},\ and\ \bibinfo {author} {\bibfnamefont {J.}~\bibnamefont
  {Behler}},\ }\bibfield  {title} {\enquote {\bibinfo {title} {Proton-transfer
  mechanisms at the water–{Z}n{O} interface: The role of presolvation},}\
  }\href {https://doi.org/10.1021/acs.jpclett.7b00358} {\bibfield  {journal}
  {\bibinfo  {journal} {J. Phys. Chem. Lett.}\ }\textbf {\bibinfo {volume}
  {8}},\ \bibinfo {pages} {1476--1483} (\bibinfo {year} {2017})}\BibitemShut
  {NoStop}%
\bibitem [{\citenamefont {Quaranta}\ \emph {et~al.}(2018)\citenamefont
  {Quaranta}, \citenamefont {Hellstr{\"o}m}, \citenamefont {Behler},
  \citenamefont {Kullgren}, \citenamefont {Mitev},\ and\ \citenamefont
  {Hermansson}}]{quaranta2018}%
  \BibitemOpen
  \bibfield  {author} {\bibinfo {author} {\bibfnamefont {V.}~\bibnamefont
  {Quaranta}}, \bibinfo {author} {\bibfnamefont {M.}~\bibnamefont
  {Hellstr{\"o}m}}, \bibinfo {author} {\bibfnamefont {J.}~\bibnamefont
  {Behler}}, \bibinfo {author} {\bibfnamefont {J.}~\bibnamefont {Kullgren}},
  \bibinfo {author} {\bibfnamefont {P.~D.}\ \bibnamefont {Mitev}},\ and\
  \bibinfo {author} {\bibfnamefont {K.}~\bibnamefont {Hermansson}},\ }\bibfield
   {title} {\enquote {\bibinfo {title} {Maximally resolved anharmonic {OH}
  vibrational spectrum of the water/{Z}n{O}(10{\=1}0) interface from a
  high-dimensional neural network potential},}\ }\href
  {https://doi.org/10.1063/1.5012980} {\bibfield  {journal} {\bibinfo
  {journal} {J. Chem. Phys.}\ }\textbf {\bibinfo {volume} {148}},\ \bibinfo
  {pages} {241720} (\bibinfo {year} {2018})}\BibitemShut {NoStop}%
\bibitem [{\citenamefont {Hellstr{\"o}m}, \citenamefont {Quaranta},\ and\
  \citenamefont {Behler}(2019)}]{hellstrom2019}%
  \BibitemOpen
  \bibfield  {author} {\bibinfo {author} {\bibfnamefont {M.}~\bibnamefont
  {Hellstr{\"o}m}}, \bibinfo {author} {\bibfnamefont {V.}~\bibnamefont
  {Quaranta}},\ and\ \bibinfo {author} {\bibfnamefont {J.}~\bibnamefont
  {Behler}},\ }\bibfield  {title} {\enquote {\bibinfo {title} {One-dimensional
  vs. two-dimensional proton transport processes at solid–liquid
  zinc-oxide–water interfaces},}\ }\href {https://doi.org/10.1039/C8SC03033B}
  {\bibfield  {journal} {\bibinfo  {journal} {Chem. Sci.}\ }\textbf {\bibinfo
  {volume} {10}},\ \bibinfo {pages} {1232--1243} (\bibinfo {year}
  {2019})}\BibitemShut {NoStop}%
\bibitem [{\citenamefont {Quaranta}, \citenamefont {Behler},\ and\
  \citenamefont {Hellstr{\"o}m}(2019)}]{quaranta2019}%
  \BibitemOpen
  \bibfield  {author} {\bibinfo {author} {\bibfnamefont {V.}~\bibnamefont
  {Quaranta}}, \bibinfo {author} {\bibfnamefont {J.}~\bibnamefont {Behler}},\
  and\ \bibinfo {author} {\bibfnamefont {M.}~\bibnamefont {Hellstr{\"o}m}},\
  }\bibfield  {title} {\enquote {\bibinfo {title} {Structure and dynamics of
  the liquid–water/zinc-oxide interface from machine learning potential
  simulations},}\ }\href {https://doi.org/10.1021/acs.jpcc.8b10781} {\bibfield
  {journal} {\bibinfo  {journal} {J. Phys. Chem. C}\ }\textbf {\bibinfo
  {volume} {123}},\ \bibinfo {pages} {1293--1304} (\bibinfo {year}
  {2019})}\BibitemShut {NoStop}%
\bibitem [{\citenamefont {Hammer}, \citenamefont {Hansen},\ and\ \citenamefont
  {N{\o}rskov}(1999)}]{hammer_improved_1999}%
  \BibitemOpen
  \bibfield  {author} {\bibinfo {author} {\bibfnamefont {B.}~\bibnamefont
  {Hammer}}, \bibinfo {author} {\bibfnamefont {L.~B.}\ \bibnamefont {Hansen}},\
  and\ \bibinfo {author} {\bibfnamefont {J.~K.}\ \bibnamefont {N{\o}rskov}},\
  }\bibfield  {title} {\enquote {\bibinfo {title} {Improved adsorption
  energetics within density-functional theory using revised
  {Perdew}-{Burke}-{Ernzerhof} functionals},}\ }\href
  {https://doi.org/10.1103/PhysRevB.59.7413} {\bibfield  {journal} {\bibinfo
  {journal} {Phys. Rev. B}\ }\textbf {\bibinfo {volume} {59}},\ \bibinfo
  {pages} {7413--7421} (\bibinfo {year} {1999})}\BibitemShut {NoStop}%
\bibitem [{\citenamefont {Grimme}\ \emph {et~al.}(2010)\citenamefont {Grimme},
  \citenamefont {Antony}, \citenamefont {Ehrlich},\ and\ \citenamefont
  {Krieg}}]{grimme_consistent_2010}%
  \BibitemOpen
  \bibfield  {author} {\bibinfo {author} {\bibfnamefont {S.}~\bibnamefont
  {Grimme}}, \bibinfo {author} {\bibfnamefont {J.}~\bibnamefont {Antony}},
  \bibinfo {author} {\bibfnamefont {S.}~\bibnamefont {Ehrlich}},\ and\ \bibinfo
  {author} {\bibfnamefont {H.}~\bibnamefont {Krieg}},\ }\bibfield  {title}
  {\enquote {\bibinfo {title} {A consistent and accurate ab initio
  parametrization of density functional dispersion correction ({DFT}-{D}) for
  the 94 elements {H}-{Pu}},}\ }\href {https://doi.org/10.1063/1.3382344}
  {\bibfield  {journal} {\bibinfo  {journal} {J. Chem. Phys.}\ }\textbf
  {\bibinfo {volume} {132}},\ \bibinfo {pages} {154104} (\bibinfo {year}
  {2010})}\BibitemShut {NoStop}%
\bibitem [{\citenamefont {Grimme}, \citenamefont {Ehrlich},\ and\ \citenamefont
  {Goerigk}(2011)}]{grimme_effect_2011}%
  \BibitemOpen
  \bibfield  {author} {\bibinfo {author} {\bibfnamefont {S.}~\bibnamefont
  {Grimme}}, \bibinfo {author} {\bibfnamefont {S.}~\bibnamefont {Ehrlich}},\
  and\ \bibinfo {author} {\bibfnamefont {L.}~\bibnamefont {Goerigk}},\
  }\bibfield  {title} {\enquote {\bibinfo {title} {Effect of the damping
  function in dispersion corrected density functional theory},}\ }\href
  {https://doi.org/https://doi.org/10.1002/jcc.21759} {\bibfield  {journal}
  {\bibinfo  {journal} {J. Comput. Chem.}\ }\textbf {\bibinfo {volume} {32}},\
  \bibinfo {pages} {1456--1465} (\bibinfo {year} {2011})}\BibitemShut {NoStop}%
\bibitem [{\citenamefont {Behler}(2024)}]{runner2018}%
  \BibitemOpen
  \bibfield  {author} {\bibinfo {author} {\bibfnamefont {J.}~\bibnamefont
  {Behler}},\ }\href@noop {} {\enquote {\bibinfo {title} {Ru{NN}er - {A} neural
  network code for high-dimensional neural network potentials},}\ }\bibinfo
  {howpublished} {Ruhr-Universit\"at Bochum} (\bibinfo {year}
  {2024})\BibitemShut {NoStop}%
\bibitem [{\citenamefont {K{\"u}hne}\ \emph {et~al.}(2020)\citenamefont
  {K{\"u}hne}, \citenamefont {Iannuzzi}, \citenamefont {Ben}, \citenamefont
  {Rybkin}, \citenamefont {Seewald}, \citenamefont {Stein}, \citenamefont
  {Laino}, \citenamefont {Khaliullin}, \citenamefont {Sch{\"u}tt},
  \citenamefont {Schiffmann}, \citenamefont {Golze}, \citenamefont {Wilhelm},
  \citenamefont {Chulkov}, \citenamefont {Bani-Hashemian}, \citenamefont
  {Weber}, \citenamefont {Bor{\v{s}}tnik}, \citenamefont {Taillefumier},
  \citenamefont {Jakobovits}, \citenamefont {Lazzaro}, \citenamefont {Pabst},
  \citenamefont {M{\"u}ller}, \citenamefont {Schade}, \citenamefont {Guidon},
  \citenamefont {Andermatt}, \citenamefont {Holmberg}, \citenamefont
  {Schenter}, \citenamefont {Hehn}, \citenamefont {Bussy}, \citenamefont
  {Belleflamme}, \citenamefont {Tabacchi}, \citenamefont {Gl{\"o}{\ss}},
  \citenamefont {Lass}, \citenamefont {Bethune}, \citenamefont {Mundy},
  \citenamefont {Plessl}, \citenamefont {Watkins}, \citenamefont
  {VandeVondele}, \citenamefont {Krack},\ and\ \citenamefont
  {Hutter}}]{kuhne_cp2k:_2020}%
  \BibitemOpen
  \bibfield  {author} {\bibinfo {author} {\bibfnamefont {T.~D.}\ \bibnamefont
  {K{\"u}hne}}, \bibinfo {author} {\bibfnamefont {M.}~\bibnamefont {Iannuzzi}},
  \bibinfo {author} {\bibfnamefont {M.~D.}\ \bibnamefont {Ben}}, \bibinfo
  {author} {\bibfnamefont {V.~V.}\ \bibnamefont {Rybkin}}, \bibinfo {author}
  {\bibfnamefont {P.}~\bibnamefont {Seewald}}, \bibinfo {author} {\bibfnamefont
  {F.}~\bibnamefont {Stein}}, \bibinfo {author} {\bibfnamefont
  {T.}~\bibnamefont {Laino}}, \bibinfo {author} {\bibfnamefont {R.~Z.}\
  \bibnamefont {Khaliullin}}, \bibinfo {author} {\bibfnamefont
  {O.}~\bibnamefont {Sch{\"u}tt}}, \bibinfo {author} {\bibfnamefont
  {F.}~\bibnamefont {Schiffmann}}, \bibinfo {author} {\bibfnamefont
  {D.}~\bibnamefont {Golze}}, \bibinfo {author} {\bibfnamefont
  {J.}~\bibnamefont {Wilhelm}}, \bibinfo {author} {\bibfnamefont
  {S.}~\bibnamefont {Chulkov}}, \bibinfo {author} {\bibfnamefont {M.~H.}\
  \bibnamefont {Bani-Hashemian}}, \bibinfo {author} {\bibfnamefont
  {V.}~\bibnamefont {Weber}}, \bibinfo {author} {\bibfnamefont
  {U.}~\bibnamefont {Bor{\v{s}}tnik}}, \bibinfo {author} {\bibfnamefont
  {M.}~\bibnamefont {Taillefumier}}, \bibinfo {author} {\bibfnamefont {A.~S.}\
  \bibnamefont {Jakobovits}}, \bibinfo {author} {\bibfnamefont
  {A.}~\bibnamefont {Lazzaro}}, \bibinfo {author} {\bibfnamefont
  {H.}~\bibnamefont {Pabst}}, \bibinfo {author} {\bibfnamefont
  {T.}~\bibnamefont {M{\"u}ller}}, \bibinfo {author} {\bibfnamefont
  {R.}~\bibnamefont {Schade}}, \bibinfo {author} {\bibfnamefont
  {M.}~\bibnamefont {Guidon}}, \bibinfo {author} {\bibfnamefont
  {S.}~\bibnamefont {Andermatt}}, \bibinfo {author} {\bibfnamefont
  {N.}~\bibnamefont {Holmberg}}, \bibinfo {author} {\bibfnamefont {G.~K.}\
  \bibnamefont {Schenter}}, \bibinfo {author} {\bibfnamefont {A.}~\bibnamefont
  {Hehn}}, \bibinfo {author} {\bibfnamefont {A.}~\bibnamefont {Bussy}},
  \bibinfo {author} {\bibfnamefont {F.}~\bibnamefont {Belleflamme}}, \bibinfo
  {author} {\bibfnamefont {G.}~\bibnamefont {Tabacchi}}, \bibinfo {author}
  {\bibfnamefont {A.}~\bibnamefont {Gl{\"o}{\ss}}}, \bibinfo {author}
  {\bibfnamefont {M.}~\bibnamefont {Lass}}, \bibinfo {author} {\bibfnamefont
  {I.}~\bibnamefont {Bethune}}, \bibinfo {author} {\bibfnamefont {C.~J.}\
  \bibnamefont {Mundy}}, \bibinfo {author} {\bibfnamefont {C.}~\bibnamefont
  {Plessl}}, \bibinfo {author} {\bibfnamefont {M.}~\bibnamefont {Watkins}},
  \bibinfo {author} {\bibfnamefont {J.}~\bibnamefont {VandeVondele}}, \bibinfo
  {author} {\bibfnamefont {M.}~\bibnamefont {Krack}},\ and\ \bibinfo {author}
  {\bibfnamefont {J.}~\bibnamefont {Hutter}},\ }\bibfield  {title} {\enquote
  {\bibinfo {title} {{CP}2k: {An} electronic structure and molecular dynamics
  software package - {Quickstep}: {Efficient} and accurate electronic structure
  calculations},}\ }\href@noop {} {\bibfield  {journal} {\bibinfo  {journal}
  {J. Chem. Phys.}\ }\textbf {\bibinfo {volume} {152}},\ \bibinfo {pages}
  {194103} (\bibinfo {year} {2020})}\BibitemShut {NoStop}%
\bibitem [{\citenamefont {Perdew}, \citenamefont {Burke},\ and\ \citenamefont
  {Ernzerhof}(1996)}]{perdew1996generalized}%
  \BibitemOpen
  \bibfield  {author} {\bibinfo {author} {\bibfnamefont {J.~P.}\ \bibnamefont
  {Perdew}}, \bibinfo {author} {\bibfnamefont {K.}~\bibnamefont {Burke}},\ and\
  \bibinfo {author} {\bibfnamefont {M.}~\bibnamefont {Ernzerhof}},\ }\bibfield
  {title} {\enquote {\bibinfo {title} {Generalized gradient approximation made
  simple},}\ }\href@noop {} {\bibfield  {journal} {\bibinfo  {journal} {Phys.
  Rev. Lett.}\ }\textbf {\bibinfo {volume} {77}},\ \bibinfo {pages} {3865}
  (\bibinfo {year} {1996})}\BibitemShut {NoStop}%
\bibitem [{\citenamefont {Zhang}\ and\ \citenamefont
  {Yang}(1998)}]{PhysRevLett.80.890}%
  \BibitemOpen
  \bibfield  {author} {\bibinfo {author} {\bibfnamefont {Y.}~\bibnamefont
  {Zhang}}\ and\ \bibinfo {author} {\bibfnamefont {W.}~\bibnamefont {Yang}},\
  }\bibfield  {title} {\enquote {\bibinfo {title} {Comment on `{G}eneralized
  gradient approximation made simple'},}\ }\href
  {https://doi.org/10.1103/PhysRevLett.80.890} {\bibfield  {journal} {\bibinfo
  {journal} {Phys. Rev. Lett.}\ }\textbf {\bibinfo {volume} {80}},\ \bibinfo
  {pages} {890--890} (\bibinfo {year} {1998})}\BibitemShut {NoStop}%
\bibitem [{\citenamefont {Adamo}\ and\ \citenamefont
  {Barone}(1999)}]{adamo1999toward}%
  \BibitemOpen
  \bibfield  {author} {\bibinfo {author} {\bibfnamefont {C.}~\bibnamefont
  {Adamo}}\ and\ \bibinfo {author} {\bibfnamefont {V.}~\bibnamefont {Barone}},\
  }\bibfield  {title} {\enquote {\bibinfo {title} {Toward reliable density
  functional methods without adjustable parameters: The {PBE}0 model},}\
  }\href@noop {} {\bibfield  {journal} {\bibinfo  {journal} {J. Chem. Phys.}\
  }\textbf {\bibinfo {volume} {110}},\ \bibinfo {pages} {6158--6170} (\bibinfo
  {year} {1999})}\BibitemShut {NoStop}%
\bibitem [{\citenamefont {Singraber}\ \emph {et~al.}(2022)\citenamefont
  {Singraber}, \citenamefont {Bircher}, \citenamefont {Reeve}, \citenamefont
  {Swenson}, \citenamefont {Lauret},\ and\ \citenamefont {David}}]{n2p2220}%
  \BibitemOpen
  \bibfield  {author} {\bibinfo {author} {\bibfnamefont {A.}~\bibnamefont
  {Singraber}}, \bibinfo {author} {\bibfnamefont {M.~P.}\ \bibnamefont
  {Bircher}}, \bibinfo {author} {\bibfnamefont {S.}~\bibnamefont {Reeve}},
  \bibinfo {author} {\bibfnamefont {D.~W.~H.}\ \bibnamefont {Swenson}},
  \bibinfo {author} {\bibfnamefont {J.}~\bibnamefont {Lauret}},\ and\ \bibinfo
  {author} {\bibfnamefont {P.}~\bibnamefont {David}},\ }\href@noop {} {\enquote
  {\bibinfo {title} {Compphysvienna/n2p2: Version 2.2.0},}\ } (\bibinfo {year}
  {2022}),\ \bibinfo {note}
  {https://github.com/CompPhysVienna/n2p2/releases/tag/v2.2.0}\BibitemShut
  {NoStop}%
\bibitem [{\citenamefont {Lobaugh}\ and\ \citenamefont
  {Voth}(1997)}]{lobaugh1997quantum}%
  \BibitemOpen
  \bibfield  {author} {\bibinfo {author} {\bibfnamefont {J.}~\bibnamefont
  {Lobaugh}}\ and\ \bibinfo {author} {\bibfnamefont {G.~A.}\ \bibnamefont
  {Voth}},\ }\bibfield  {title} {\enquote {\bibinfo {title} {A quantum model
  for water: Equilibrium and dynamical properties},}\ }\href@noop {} {\bibfield
   {journal} {\bibinfo  {journal} {J. Chem. Phys.}\ }\textbf {\bibinfo {volume}
  {106}},\ \bibinfo {pages} {2400--2410} (\bibinfo {year} {1997})}\BibitemShut
  {NoStop}%
\bibitem [{\citenamefont {Shiga}(2025)}]{shiga2025pimd}%
  \BibitemOpen
  \bibfield  {author} {\bibinfo {author} {\bibfnamefont {M.}~\bibnamefont
  {Shiga}},\ }\href@noop {} {\enquote {\bibinfo {title} {{PIMD}: An open-source
  software for parallel molecular simulations},}\ } (\bibinfo {year} {2025}),\
  \bibinfo {note}
  {https://ccse.jaea.go.jp/software/PIMD/index.en.html}\BibitemShut {NoStop}%
\bibitem [{\citenamefont {Soper}(2013)}]{soper_radial_2013}%
  \BibitemOpen
  \bibfield  {author} {\bibinfo {author} {\bibfnamefont {A.~K.}\ \bibnamefont
  {Soper}},\ }\bibfield  {title} {\enquote {\bibinfo {title} {The radial
  distribution functions of water as derived from radiation total scattering
  experiments: {I}s there anything we can say for sure?}}\ }\href@noop {}
  {\bibfield  {journal} {\bibinfo  {journal} {ISRN Physical Chemistry}\
  }\textbf {\bibinfo {volume} {2013}},\ \bibinfo {pages} {e279463} (\bibinfo
  {year} {2013})}\BibitemShut {NoStop}%
\bibitem [{sop()}]{soper_radial_2000}%
  \BibitemOpen
  \bibfield  {title} {\enquote {\bibinfo {title} {The radial distribution
  functions of water and ice from 220 to 673 {K} and at pressures up to 400
  {M}pa, volume = {258}, issn = {0301-0104}, language = {en}, number = {2},
  urldate = {2021-04-28}, journal = {Chem. Phys.}, author = {Soper, A. K.},
  month = aug, year = {2000}, pages = {121--137},},}\ }\href@noop {} {\
  }\BibitemShut {NoStop}%
\bibitem [{\citenamefont {Machida}, \citenamefont {Kato},\ and\ \citenamefont
  {Shiga}(2018)}]{machida2018nuclear}%
  \BibitemOpen
  \bibfield  {author} {\bibinfo {author} {\bibfnamefont {M.}~\bibnamefont
  {Machida}}, \bibinfo {author} {\bibfnamefont {K.}~\bibnamefont {Kato}},\ and\
  \bibinfo {author} {\bibfnamefont {M.}~\bibnamefont {Shiga}},\ }\bibfield
  {title} {\enquote {\bibinfo {title} {Nuclear quantum effects of light and
  heavy water studied by all-electron first principles path integral
  simulations},}\ }\href@noop {} {\bibfield  {journal} {\bibinfo  {journal} {J.
  Chem. Phys.}\ }\textbf {\bibinfo {volume} {148}},\ \bibinfo {pages} {102324}
  (\bibinfo {year} {2018})}\BibitemShut {NoStop}%
\bibitem [{\citenamefont {Wagner}\ and\ \citenamefont
  {Pru{\ss}}(2002)}]{IAPWS1995}%
  \BibitemOpen
  \bibfield  {author} {\bibinfo {author} {\bibfnamefont {W.}~\bibnamefont
  {Wagner}}\ and\ \bibinfo {author} {\bibfnamefont {A.}~\bibnamefont
  {Pru{\ss}}},\ }\bibfield  {title} {\enquote {\bibinfo {title} {The {IAPWS}
  formulation 1995 for the thermodynamic properties of ordinary water substance
  for general and scientific use},}\ }\href {https://doi.org/10.1063/1.1461829}
  {\bibfield  {journal} {\bibinfo  {journal} {J. Phys. Chem. Ref. Data}\
  }\textbf {\bibinfo {volume} {31}},\ \bibinfo {pages} {387--535} (\bibinfo
  {year} {2002})}\BibitemShut {NoStop}%
\bibitem [{\citenamefont {Flyvbjerg}\ and\ \citenamefont
  {Petersen}(1989)}]{flyvbjerg1989error}%
  \BibitemOpen
  \bibfield  {author} {\bibinfo {author} {\bibfnamefont {H.}~\bibnamefont
  {Flyvbjerg}}\ and\ \bibinfo {author} {\bibfnamefont {H.~G.}\ \bibnamefont
  {Petersen}},\ }\bibfield  {title} {\enquote {\bibinfo {title} {Error
  estimates on averages of correlated data},}\ }\href@noop {} {\bibfield
  {journal} {\bibinfo  {journal} {J. Chem. Phys.}\ }\textbf {\bibinfo {volume}
  {91}},\ \bibinfo {pages} {461--466} (\bibinfo {year} {1989})}\BibitemShut
  {NoStop}%
\end{thebibliography}%
%%%%%%%%%%%%%%%%%%%%%%%%%%%%%%%%%%%%%%%%%%%%%%%%%%%%%%%%%%%%%%%%%%%%%%%%

%%%%%%%%%%%%%%%%%%%%%%%%%%%%%%%%%%%%%%%%%%%%%%%%%%%%%%%%%%%%%%%%%%%%%%%%
\end{document}